\documentclass[preprint,12pt]{elsarticle}

\usepackage{graphics}

\journal{Physica A}

\begin{document}

\begin{frontmatter}

\title{Spin-glass transition in the Ising model \\
on multiplex networks}

\author{A. Krawiecki}       

\address{Faculty of Physics,
Warsaw University of Technology, \\
Koszykowa 75, PL-00-662 Warsaw, Poland}

\begin{abstract}
Multiplex networks consist of a fixed set of nodes connected by several sets of 
edges which are generated separately and correspond to different networks ("layers"). 
Here, the Ising model on multiplex networks with two layers is considered, with spins
located in the nodes and edges corresponding to ferromagnetic or antiferromagnetic interactions between them.
Critical temperatures for the spin glass and ferromagnetic transitions are evaluated for the layers in the form of random
Erd\"os-R\'enyi graphs or heterogeneous scale-free networks using the replica method, from the replica symmetric solution. 
For the Ising model on multiplex networks with
scale-free layers it is shown that the critical temperature is finite if the distributions of the degrees of nodes within both layers have a finite
second moment, and that depending on the model parameters the transition can be to the ferromagnetic or spin glass phase. It is also shown that the 
correlation between the degrees of nodes within different layers significantly influences the critical temperatures for both transitions
and thus the phase diagram. The scaling behavior for the spin glass order parameter is determined by a semi-analytic procedure
and it is shown that for the Ising model on multiplex networks with scale-free layers the scaling exponent can depend on the 
distributions of the degrees of nodes within layers. The analytic results are partly confirmed by Monte Carlo simulations using
the parallel tempering algorithm.
\end{abstract}


\end{frontmatter}

\section{Introduction}

In the last two decades rapid advancement in the theory and applications of complex networks has taken place
related to the widespread recognition of their importance in social life, natural sciences and technology \cite{Albert02,Barabasi16}.
An important part of this trend was development of research on complex systems in which interactions among their
constituent parts are determined by the underlying structure of complex networks \cite{Dorogovtsev08,Barrat08}.
In this context much effort was devoted to study the effect of the complex structure of interactions on the behavior of generic models 
of statistical physics exhibiting collective phenomena such as phase transitions. For example, ferromagnetic (FM) phase transition
in the Ising model on complex, possibly heterogeneous networks was studied by means of various analytic 
\cite{Bianconi02,Leone02,Dorogovtsev02,Yoon11} and numerical \cite{Herrero04,Herrero15} methods.
Also spin glass (SG) transition \cite{Mezard87,Nishimori01} in the Ising and related models on complex networks with quenched disorder
of FM and antiferromagnetic (AFM) interactions was investigated using, e.g., variants of the replica method
\cite{Nikoletopoulos04,Wemmenhove05,Kim05,Kim14}, effective field theory \cite{Ostilli08,Ferreira10} and Monte Carlo (MC)
simulations \cite{Bartolozzi06,Herrero09}. In connection with recent interest in even more complex structures ("networks of networks")
much attention has been devoted to multiplex networks (MNs) which consist of a fixed set of nodes connected by various
sets of edges called layers \cite{Boccaletti14,Lee14,Lee15}. MNs naturally emerge in many social systems (e.g., transportation or communications networks),
and interacting systems on such structures exhibit rich variety of collective behaviors and critical phenomena. For example,
percolation transition \cite{Buldyrev10,Baxter12,Min14}, cascading failures \cite{Tan13}, diffusion processes \cite{Gomez13,Sole13}, 
epidemic spreading \cite{Wu16,Zuzek15}, etc., were studied on MNs. Also the FM transition in the Ising model \cite{Krawiecki17} 
as well as diversity of first-order, second order and mixed-order transitions in a related Ashkin-Teller model \cite{Jang15} were investigated
in the above-mentioned models with the structure of MNs.

As a natural extension of the above-mentioned research in this paper the SG transition is studied in the Ising model with the 
quenched disorder of the exchange interactions superimposed on the underlying structure of a MN. In Sec.\ 2 the Hamiltonian of the model
is defined, with spins placed on a fixed set of nodes and with separately generated sets of edges (layers), with possibly different 
distributions of the degrees of nodes, corresponding to randomly assigned
FM and AFM exchange interactions; the layers can have, e.g., the structure of random 
Erd\"os-R\'enyi (ER) graphs \cite{Erdos59} or heterogeneous scale-free (SF) networks \cite{Barabasi99} and are generated from the so-called
static model \cite{Goh01,Lee04}. In Sec.\ 3 the thermodynamic properties of the above-mentioned model are investigated by means of the 
replica method \cite{Mezard87,Nishimori01}. The approach used here follows the study of the dilute SG model with infinite-range interactions
\cite{Viana85,Kanter87,Mezard87a,Mottishaw87,Monasson98,Castellani05,Hase12} and is a direct generalization to the case of MNs
of a procedure applied successfully to investigate the FM and SG transitions in the Ising model on random ER graphs \cite{Viana85},
heterogeneous SF networks \cite{Kim05} and the FM transition in the Ising model on MNs \cite{Krawiecki17}.
In Sec.\ 4 the FM and SG transitions from the paramagnetic state are investigated in the above-mentioned model,
the corresponding critical temperatures are evaluated from the replica symmetric (RS) solution and the effect of the distributions of the degrees of nodes within
consecutive layers as well as the influence of the correlations between them on the phase diagram is emphasised. Besides, these analytic results are
partly compared with MC simulations. In Sec.\ 5 the critical exponent for the SG order parameter in the vicinity of the SG transition temperature 
is determined semi-analytically for the Ising model on MNs with different distributions of the degrees of nodes within layers. 
Sec.\ 6 is devoted to summary and conclusions.

\section{The model}

\subsection{The Hamiltonian}

MNs consist of a fixed set of nodes connected by several sets of edges; the set of nodes with each set of edges forms a
network which is called a layer of a MN \cite{Lee14,Lee15}. In this paper only fully overlapping
MNs are considered, with all $N$ nodes belonging to all layers. In the following, for simplicity, MNs with $N$ nodes and only two
layers denoted as $G^{(A)}$, $G^{(B)}$ are considered. The layers (strictly speaking, the sets of edges of each layer)
are generated separately, and, possibly, independently. As a result, multiple connections between nodes are not
allowed within the same layer, but the same nodes can be connected by multiple edges belonging to different layers.
The nodes $i=1,2,\ldots N$ are characterized by their degrees $k_{i}^{(A)}$,
$k_{i}^{(B)}$ within each layer, i.e., the number of edges attached to them within each layer. The, possibly heterogeneous,
distributions of the degrees of nodes within each layer are denoted as $p_{k^{(A)}}$, $p_{k^{(B)}}$, and the
mean degrees of nodes within each layer as $\langle k^{(A)} \rangle$, $\langle k^{(B)} \rangle$.

In the Ising model on a MN with two layers two-state spins $s_{i}=\pm 1$ are located in the nodes $i=1,2\ldots N$ and edges
within the layers $G^{(A)}$, $G^{(B)}$ connecting pairs of nodes $i$, $j$ 
correspond to exchange interactions with integrals $J_{ij}^{(A)}$, $J_{ij}^{(B)}$, respectively.
The exchange integrals are quenched random variables.
It should be emphasised that in the model under study there is only one spin $s_{i}$ located in each node which interacts with all its neighbors
within all layers. The Hamiltonian of the model is
\begin{equation}
H= - \sum_{\left( i,j\right) \in G^{(A)}}  J_{ij}^{(A)}s_{i}s_{j} -  \sum_{\left( i,j\right) \in G^{(B)}}  J_{ij}^{(B)} s_{i}s_{j},
\label{ham}
\end{equation}
where the sums are over all edges belonging to the layer $G^{(A)}$ ($G^{(B)}$). 

Following the studies of the dilute Ising SG models with
infinite-range interactions on random ER graphs \cite{Viana85} and SF networks \cite{Kim05} in this paper it is assumed that
the exchange integrals within each layer can assume only two values $J^{(A)}$ ($J^{(A)}>0$) and $-J^{(A)}$ 
as well as $J^{(B)}$ ($J^{(B)}>0$) and $-J^{(B)}$ which are assigned to the edges of the layer $G^{(A)}$ ($G^{(B)}$)
with probability $r^{(A)}$ and $1-r^{(A)}$ ($r^{(B)}$ and $1-r^{(B)}$), respectively, and that these assignments are independent for the
two layers. Thus the distributions of the exchange integrals within each layer $P_{r^{(A)}}\left(\left\{ J_{ij}^{(A)}\right\} \right)$,
$P_{r^{(B)}}\left( \left\{ J_{ij}^{(B)} \right\} \right)$ are independent and have the form
\begin{eqnarray}
P_{r^{(A)}}\left( \left\{ J_{ij}^{(A)}\right\} \right) &=& \prod_{\left( i,j\right) \in G^{(A)}} \left[ r^{(A)} \delta\left(  J_{ij}^{(A)} -J^{(A)}\right) 
+\left( 1-r^{(A)}\right) \delta \left(  J_{ij}^{(A)} +J^{(A)}\right) \right] \nonumber\\
P_{r^{(B)}}\left(\left\{ J_{ij}^{(B)} \right\} \right) &=& \prod_{\left( i,j\right) \in G^{(B)}} \left[ r^{(B)} \delta\left(  J_{ij}^{(B)} -J^{(B)}\right) 
+\left( 1-r^{(B)}\right) \delta \left(  J_{ij}^{(B)} +J^{(B)}\right) \right]. \nonumber\\
&& \label{pApB}
\end{eqnarray}

Taking into account the form of the Hamiltonian, Eq.\ (\ref{ham}), it may be supposed that the SG Ising model on a MN can be reduced to
the SG Ising model on a network with a set of edges being a superposition of the sets of edges of the two layers and a proper four-point
distribution of the exchange integrals $J_{ij}$. However, these two Ising models are not equivalent to each other since in a MN the layers
are generated separately (although not necessariliy completely independently) and thus, e.g., probabilities for the pairs of nodes to be connected by
an edge or statistical averages over different realizations of the sets of edges should be evaluated separately for each layer. 
This difference is particularly imporatnt in the case of MNs with heterogeneous layers, where
it was shown in Ref.\ \cite{Krawiecki17} that even in the simplest case of purely FM interactions 
with $J^{(A)}=J^{(B)}$ the critical temperatures for the two above-mentioned
Ising models can differ noticeably.
 
\subsection{The multiplex network model}

In Ref.\ \cite{Kim05} SG transition was investigated in the Ising model on heterogeneous networks generated from the static model \cite{Goh01,Lee04}.
Using this model networks with a fixed number of nodes $N$ and desired distributions of the degrees of nodes can be generated as follows. 
First, a weight $v_{i}$ is assigned to each node so that the condition $\sum_{i=1}^{N} v_{i}=1$ was fulfilled. Then, 
nodes are linked with edges in accordance with the prescribed sequence of weights, by selecting
a pair of nodes $i$, $j$ ($i\neq j$) with probablities $v_{i}$, $v_{j}$, respectively, linking them with an edge and repeating this process $NK/2$ times. In this way
a network is obtained with the probability that the nodes $i$, $j$ are linked by an edge $f_{ij}\approx NK v_{i}v_{j}$, with the mean degree of nodes 
$\langle k\rangle =K$, and with the distribution of the degrees of nodes depending on the choice of the weights. 
In particular, random ER graph is obtained if $v_{i}=1/N$ is assumed for all $i$. 
For a sequence $v_{i}=i^{-\mu}/\zeta_{N}( \mu)$ associated with the nodes $i=1,2,\ldots N$, where $0< \mu <1$ and 
$\zeta_{N}( \mu ) \approx N^{1-\mu}/(1-\mu)$,
SF network is obtained with the distribution of the degrees of nodes $p_{k}\propto k^{-\gamma}$, $\gamma = 1 +1/\mu$. 
In an ensemble of networks generated from the static model the mean degree of a given node $i$ is $\langle k_{i}\rangle =NK v_{i}$. 

Similarly, in this paper the Ising model on a MN with layers generated from the static model is studied. 
The MN with a fixed set of nodes and two layers $G^{(A)}$, $G^{(B)}$ is generated by associating weights $v_{i}^{(A)}$, $v_{i}^{(B)}$ with the nodes 
separately to generate each layer. In this way the layers can have different distributions of the degerees of nodes $p_{k^{(A)}}$, $p_{k^{(B)}}$.
Let us note that
the numbering of nodes $i=1,2,\ldots N$ while generating each layer can be assumed the same or different. In the case of random ER layers this distinction is unimportant, however,
in the case of SF layers it can introduce correlations between the two sequences of weights $v_{i}^{(A)}$, $v_{i}^{(B)}$, $i=1,2,\ldots N$, where now and 
henceforth $i$ denotes the index of the node in a MN, common for all layers. In particular, a MN with independent
layers is obtained by randomly and independently associating weights from the two appropriate sets of weights with the nodes and then linking them with
edges according to the prescribed sequence of weights within each layer.

\section{Evaluation of the free energy using the replica method}

\subsection{General considerations}

The starting point to study the thermodynamic properties of the Ising model on MNs is to evaluate the free energy 
averaged over a statistical ensemble of 
MNs generated according to a given rule and with given quenched disorder of the exchange integrals. Hence, the free energy is 
$-\beta F=\left[ \left[ \ln Z\right]_{r}\right]_{av}$, 
where $Z$ is the partition function for the Ising model on a particular MN with a particular distribution of $J_{ij}$, 
the average $\left[ \cdot \right]_{av}$ is taken over all possible random realizations of a set of edges in a MN of a given kind 
(i.e., of the two sets of edges in the separately generated layers), 
and the average $\left[ \cdot \right]_{r}$ is taken over all possible realizations of the distributions
$P_{r^{(A)}} \left( \left\{ J_{ij}^{(A)} \right\}\right)$, $P_{r^{(B)}} \left( \left\{ J_{ij}^{(B)} \right\}\right)$, Eq.\ (\ref{pApB}),
for fixed sets of edges within each layer $G^{(A)}$, $G^{(B)}$. In the framework of the replica method the free energy is formally evaluated as
$-\beta F =\lim_{n\rightarrow 0} \left\{ \left[\left[  Z^{n}\right]_{r}\right]_{av} -1\right\}/n$. The 
average of the $n$-th power of the partition function is
\begin{equation}
\left[\left[  Z^{n}\right]_{r} \right]_{av} = {\rm Tr}_{\{s^{\alpha}\}}
\left[ \left[ \exp\left( \beta \sum_{\left( i,j\right)\in G^{(A)}} J_{ij}^{(A)} \sum_{\alpha =1}^{n}s_{i}^{\alpha}s_{j}^{\alpha}\right)
\exp\left( \beta \sum_{\left( i,j\right)\in G^{(B)}} J_{ij}^{(B)} \sum_{\alpha =1}^{n}s_{i}^{\alpha}s_{j}^{\alpha}\right) \right]_{r}\right]_{av},
\label{Zn}
\end{equation}
i.e., it is the average of a product of $n$ partition functions for non-interacting replicas (copies) of the system,
the trace ${\rm Tr}_{\left\{ s^{\alpha}\right\}}$ is taken over all replicated spins $s_{i}^{\alpha} =\pm 1$, and $\alpha =1,2\ldots n$ is the replica
index.

As pointed out in Ref.\ \cite{Krawiecki17} generation of a MN takes place in two stages: first, in which the weights
$v_{i}^{(A)}$, $v_{i}^{(B)}$ are separately assigned to the nodes $i=1,2,\ldots N$, and second, in which the nodes are connected
with edges taking into account the prescribed weights within each layer. At the first stage the weights from the two sets of weights can be assigned to the
nodes either independently or certain correlations between the two weights assigned to the same nodes can be present (e.g., 
higher weights from both sets can be assigned to the same nodes). Such correlations can change substantially the thermodynamic
properties of the model, e.g., the critical temperature for the FM transition \cite{Krawiecki17}.
Thus, it is necessary to consider separately classes of MNs characterized
by given pairs of sequences of weights $v_{i}^{(A)}$, $v_{i}^{(B)}$,  $i=1,2,\ldots N$. Then the average $\left[ \cdot \right]_{av}$ in
Eq.\ (\ref{Zn}) is evaluated separately for each class, and is taken over all possible realizations of the two layers by connecting the
nodes with edges according to the weights $v_{i}^{(A)}$, $v_{i}^{(B)}$, $i=1,2,\ldots N$ characterizing this class.
If necessary, a sort of further averaging over different classes of MNs (e.g., over all classes with the same 
correlation coefficient between the two sequences of weights $v_{i}^{(A)}$, $v_{i}^{(B)}$, $i=1,2,\ldots N$) can be
performed by replacing the sums over $N$ nodes by their expected values in the resulting expressions for the critical 
temperature.

For a class of MNs with fixed (correlated or not) assignment of the weights $v_{i}^{(A)}$, $v_{i}^{(B)}$ to the nodes 
the sets of edges of each layer are generated independently of each other. Thus
the average over all realizations of the set of edges of a MN in Eq.\ (\ref{Zn})
can be taken independently over all realizations of the sets of edges in the layers $G^{(A)}$ and $G^{(B)}$ in accordance with these weights.
Denoting the respective averages by $\left[ \cdot \right]_{av}^{(A)}$, $\left[ \cdot \right]_{av}^{(B)}$, taking into account that
assignment of the exchange integrals $J_{ij}^{(A)}$, $J_{ij}^{(B)}$ to the edges of the layers $G^{(A)}$, $G^{(B)}$ also takes place
independently for each layer and denoting the averages over all possible realizations of the distributions
$P_{r^{(A)}} \left( \left\{ J_{ij}^{(A)} \right\}\right)$ and $P_{r^{(B)}} \left( \left\{ J_{ij}^{(B)} \right\}\right)$ as
$\left[ \cdot \right]_{r^{(A)}}$ and $\left[ \cdot \right]_{r^{(B)}}$, respectively, Eq.\ (\ref{Zn}) can be written as
\begin{eqnarray}
\left[ \left[ Z^{n}\right]_{r} \right]_{av}&=& 
{\rm Tr}_{\{s^{\alpha}\}} \left\{
\left[ \left[ \exp\left( \beta \sum_{\left( i,j\right)\in G^{(A)}} J_{ij}^{(A)}\sum_{\alpha =1}^{n}s_{i}^{\alpha}s_{j}^{\alpha}\right) \right]_{r^{(A)}} \right]_{av}^{(A)} \right. \nonumber\\
&\times& \left. \left[ \left[ \exp\left( \beta \sum_{\left( i,j\right)\in G^{(B)}} J_{ij}^{(B)}\sum_{\alpha =1}^{n}s_{i}^{\alpha}s_{j}^{\alpha}\right)  \right]_{r^{(B)}} \right]_{av}^{(B)}\right\}.
\label{Zn1}
\end{eqnarray}
The two factors can be evaluated as in Ref.\ \cite{Kim05},
\begin{eqnarray}
&& \left[ \left[
\exp\left( \beta \sum_{\left( i,j\right)\in G^{(A)}} J_{ij}^{(A)} \sum_{\alpha =1}^{n}s_{i}^{\alpha}s_{j}^{\alpha}\right) 
\right]_{r^{(A)}} \right]_{av}^{(A)} =  \nonumber\\
&& \prod_{i<j} \left\{ \left( 1-f_{ij}^{(A)} \right) + f_{ij}^{(A)} 
\left[ \exp\left( \beta J_{ij}^{(A)} \sum_{\alpha =1}^{n}s_{i}^{\alpha}s_{j}^{\alpha}\right) \right]_{r^{(A)}} \right\}  = \nonumber\\
&& \exp \left\{ \sum_{i<j} \ln \left[ 1+f_{ij}^{(A)} 
\left[ \exp\left( \beta J_{ij}^{(A)}  \sum_{\alpha =1}^{n}s_{i}^{\alpha}s_{j}^{\alpha}\right) -1\right]_{r^{(A)}} \right] \right\} 
\approx \nonumber\\
&& \exp\left[ \sum_{i<j} NK^{(A)} v_{i}^{(A)}v_{j}^{(A)}\left[ \exp\left( \beta J_{ij}^{(A)} \sum_{\alpha =1}^{n}s_{i}^{\alpha}s_{j}^{\alpha}\right) -1
\right]_{r^{(A)}} \right],
\label{Zn2}
\end{eqnarray}
and similarly for the average $\left[ \left[ \cdot \right]_{r^{(B)}} \right]_{av}^{(B)}$. Then, since $s_{i}^{\alpha}s_{j}^{\alpha} = \pm 1$, the relation
\begin{equation}
\left[ \exp\left( \beta J_{ij}^{(A)} \sum_{\alpha =1}^{n}s_{i}^{\alpha}s_{j}^{\alpha}\right) \right]_{r^{(A)}}
= \left[ \prod_{\alpha} \cosh \beta J^{(A)} \left( 1+ s_{i}^{\alpha}s_{j}^{\alpha} \tanh\beta J^{(A)} \right) \right]_{r^{(A)}}
\end{equation}
can be used in Eq.\ (\ref{Zn2}), which yields
\begin{eqnarray}
&& \left[ \left[ 
\exp\left( \beta \sum_{\left( i,j\right)\in G^{(A)}} J_{ij}^{(A)}\sum_{\alpha =1}^{n}s_{i}^{\alpha}s_{j}^{\alpha}\right)
\right]_{r^{(A)}} \right]_{av}^{(A)} \propto  \nonumber\\
&& \exp\left[ \sum_{i<j} NK^{(A)} v_{i}^{(A)}v_{j}^{(A)} \left( {\bf T}_{1}^{(A)} \sum_{\alpha} s_{i}^{\alpha}s_{j}^{\alpha} +
{\bf T}_{2}^{(A)} \sum_{\alpha < \beta} s_{i}^{\alpha}s_{i}^{\beta} s_{j}^{\alpha}s_{j}^{\beta} +\ldots \right) \right],
\label{Zn4}
\end{eqnarray}
where 
\begin{eqnarray}
{\bf T}_{1}^{(A)} &=& \left[ \cosh^{n}\beta J_{ij}^{(A)} \tanh\beta J_{ij}^{(A)} \right]_{r^{(A)}}
 \stackrel{n\rightarrow 0}{\rightarrow} (2r^{(A)}-1) \tanh \beta J^{(A)}, \nonumber\\
{\bf T}_{2}^{(A)}&=& \left[ \cosh^{n}\beta J_{ij}^{(A)} \tanh^{2}\beta J_{ij}^{(A)}\right]_{r^{(A)}}
 \stackrel{n\rightarrow 0}{\rightarrow} \tanh^{2} \beta J^{(A)}, 
\label{T1T2}
\end{eqnarray}
etc.; similar expansion can be obtained for the average $\left[ \left[\cdot \right]_{r^{(B)}}\right]_{av}^{(B)}$.
Finally, after applying the Hubbard-Stratonovich identity to the expressions of the form (\ref{Zn4}), separately for the two averages 
 $\left[ \left[\cdot \right]_{r^{(A)}}\right]_{av}^{(A)}$,  $\left[ \left[\cdot \right]_{r^{(B)}}\right]_{av}^{(B)}$, 
and grouping terms connected with the same nodes $i$ it is obtained that
\begin{eqnarray}
&& \left[ \left[ Z^{n}\right]_{r}\right]_{av} = \nonumber \\
&& \int dq_{\alpha}^{(A)} \int dq_{\alpha \beta}^{(A)}\ldots \int dq_{\alpha}^{(B)} \int dq_{\alpha\beta }^{(B)}\ldots 
\exp \left[ -Nn\beta f \left( q_{\alpha}^{(A)}, q_{\alpha\beta}^{(A)},\ldots q_{\alpha}^{(B)}, q_{\alpha\beta }^{(B)} \ldots \right) \right] \nonumber \\
&& \equiv \int \ d{\bf q} \exp \left[ -Nn\beta f({\bf q}) \right],
\label{Zn6}
\end{eqnarray}
with
\begin{eqnarray}
n\beta  f({\bf q})& =& \frac{K^{(A)} {\bf T}_{1}^{(A)}}{2} \sum_{\alpha} q_{\alpha}^{(A)2} 
+ \frac{K^{(B)} {\bf T}_{1}^{(B)}}{2} \sum_{\alpha} q_{\alpha}^{(B)2} 
+\ldots \nonumber\\
&+& \frac{K^{(A)} {\bf T}_{2}^{(A)}}{2} \sum_{\alpha < \beta} q_{\alpha\beta}^{(A)2} + 
\frac{K^{(B)} {\bf T}_{2}^{(B)}}{2} \sum_{\alpha< \beta} q_{\alpha\beta}^{(B)2} +\ldots \nonumber \\
&-& \frac{1}{N} \sum_{i} \ln {\rm Tr}_{\left\{ s_{i}^{\alpha}\right\}} \exp\left(  X_{i}^{(A)} +X_{i}^{(B)} \right),
\label{Zn5}
\end{eqnarray}
where ${\rm Tr}_{\left\{ s_{i}^{\alpha}\right\}}$ is the trace over the replicated spins at node $i$, and
\begin{equation}
X_{i}^{(A)} = NK^{(A)}{\bf T}_{1}^{(A)} v_{i}^{(A)} \sum_{\alpha}q_{\alpha}^{(A)} s_{i}^{\alpha} +
 NK^{(A)}{\bf T}_{2}^{(A)} v_{i}^{(A)} \sum_{\alpha < \beta}q_{\alpha\beta}^{(A)} s_{i}^{\alpha} s_{i}^{\beta} +\ldots,
\end{equation}
and similarily for $X_{i}^{(B)}$. 

The elements of a set $\{\bf q\}$, $q_{\alpha}^{(A)}, q_{\alpha\beta}^{(A)}, \ldots, q_{\alpha}^{(B)}, q_{\alpha\beta}^{(B)},
\ldots $ form in a natural way two subsets of the order parameters associated with the two layers of the multiplex network
$G_{A}$, $G_{B}$. The first two order parameters,
\begin{equation}
q_{\alpha}^{(A)} = \sum_{i}v_{i}^{(A)} \overline{s_{i}^{\alpha}}, \; \; q_{\alpha}^{(B)} = \sum_{i}v_{i}^{(B)} \overline{s_{i}^{\alpha}}, 
\label{ordpar}
\end{equation}
where the averages are evaluated as 
\begin{displaymath}
\overline{s_{i}^{\alpha}} = \frac{{\rm Tr}_{\left\{ s_{i}^{\alpha}\right\}} s_{i}^{\alpha}  \exp\left(  X_{i}^{(A)} +X_{i}^{(B)} \right)}
{{\rm Tr}_{\left\{ s_{i}^{\alpha}\right\}}  \exp\left(  X_{i}^{(A)} +X_{i}^{(B)} \right)},
\end{displaymath}
are called magnetizations for convenience; the next two order parameters 
\begin{equation}
q_{\alpha\beta}^{(A)} = \sum_{i}v_{i}^{(A)} \overline{s_{i}^{\alpha} s_{i}^{\beta}}, \; \; 
q_{\alpha\beta}^{(B)} = \sum_{i}v_{i}^{(B)} \overline{s_{i}^{\alpha} s_{i}^{\beta}}, 
\label{ordpar1}
\end{equation}
are called SG order parameters, etc.

\subsection{The replica symmetric free energy}

The simplest RS solution for the order parameters is obtained under the assumption that spins with different replica index
are indistinguishable. In the case of the Ising model on a MN this solution has a form
$q_{\alpha}^{(A)} =m^{(A)}$, $q_{\alpha \beta}^{(A)}=q^{(A)}$, etc., and
$q_{\alpha}^{(B)} =m^{(B)}$, $q_{\alpha \beta}^{(B)}=q^{(B)}$, etc., for $\alpha, \beta =1,2\ldots n$, etc., where, in general,
$m^{(A)}\neq m^{(B)}$, $q^{(A)}\neq q^{(B)}$, etc.\ \cite{Krawiecki17}.

Assuming the above-mentioned form of the RS solution and truncating the free energy, Eq.\ (\ref{Zn5}), at the order of $q^{2}$ yields
\begin{eqnarray}
&& n\beta f\left( m^{(A)},m^{(B)},q^{(A)},q^{(B)}\right)= \nonumber\\ 
&&\frac{K^{(A)} {\bf T}_{1}^{(A)}}{2}nm^{(A)2}+\frac{K^{(B)} {\bf T}_{1}^{(B)}}{2}nm^{(B)2} +\nonumber\\
&&\frac{K^{(A)} {\bf T}_{2}^{(A)}}{2} \frac{n(n-1)}{2}q^{(A)2} + \frac{K^{(B)} {\bf T}_{2}^{(B)}}{2} \frac{n(n-1)}{2}q^{(B)2} - \frac{1}{N} \sum_{i} \ln {\cal Z}_{i},
\label{nBf}
\end{eqnarray}
where
\begin{eqnarray}
{\cal Z}_{i}&=& {\rm Tr}_{\left\{ s_{i}^{\alpha}\right\} }\exp \left[ N \left(  K^{(A)}{\bf T}_{1}^{(A)}  v_{i}^{(A)} m^{(A)} + 
K^{(B)} {\bf T}_{1}^{(B)}  v_{i}^{(B)} m^{(B)} \right)
\sum_{\alpha} s_{i}^{\alpha} + \right. \nonumber\\
&& \left.  N \left(  K^{(A)} {\bf T}_{2}^{(A)} v_{i}^{(A)} q^{(A)} +  K^{(B)}{\bf T}_{2}^{(B)} v_{i}^{(B)} q^{(B)} \right)
\frac{\left( \sum_{\alpha} s_{i}^{\alpha} \right)^{2}-n}{2} \right].
\end{eqnarray}
Using once more the Hubbard-Stratonovich identity and evaluating the trace over $s_{i}^{\alpha}=\pm1$ yields
\begin{equation}
{\cal Z}_{i}= \exp\left[ -\frac{nN}{2}\left( K^{(A)}{\bf T}_{2}^{(A)} v_{i}^{(A)} q^{(A)} +  K^{(B)}{\bf T}_{2}^{(B)} v_{i}^{(B)} q^{(B)} \right)\right]
\frac{1}{\sqrt{2\pi}} \int dz e^{-z^{2}/2}\left[2 \cosh \eta_{i}(z) \right]^{n},
\label{Zi}
\end{equation}
where
\begin{eqnarray}
\eta_{i}(z) &=& N  \left( K^{(A)}{\bf T}_{1}^{(A)} v_{i}^{(A)} m^{(A)} + K^{(B)} {\bf T}_{1}^{(B)} v_{i}^{(B)} m^{(B)} \right) + \nonumber\\
&& z \sqrt{  N  \left( K^{(A)}{\bf T}_{2}^{(A)} v_{i}^{(A)} q^{(A)} + K^{(B)} {\bf T}_{2}^{(B)} v_{i}^{(B)} q^{(B)} \right) }.
\end{eqnarray}
In the limit $n\rightarrow 0$ it is possible to expand $\left[ 2\cosh \eta_{i}(z)\right]^{n} \approx 1+n \ln \left[ 2\cosh \eta_{i}(z)\right]$. 
Inserting this expansion in Eq.\ (\ref{Zi}) and again expanding $\ln {\cal Z}_{i}$ in Eq.\ (\ref{nBf}) yields for $n\rightarrow 0$
\begin{eqnarray}
&& \beta f\left( m^{(A)},m^{(B)},q^{(A)},q^{(B)}\right)= \nonumber\\
&& \frac{K^{(A)} {\bf T}_{1}^{(A)}}{2}m^{(A)2}+\frac{K^{(B)} {\bf T}_{1}^{(B)}}{2}m^{(B)2} +
\frac{K^{(A)} {\bf T}_{2}^{(A)}}{2}q^{(A)}+\frac{K^{(B)} {\bf T}_{2}^{(B)}}{2}q^{(B)} - \nonumber\\
&& \frac{K^{(A)} {\bf T}_{2}^{(A)}}{4}q^{(A)2}-\frac{K^{(B)} {\bf T}_{2}^{(B)}}{4}q^{(B)2} - 
\frac{1}{\sqrt{2\pi}} \int dz e^{-z^{2}/2}\frac{1}{N}\sum_{i} \ln \left[2 \cosh \eta_{i}(z) \right].\nonumber\\
&&
\label{Bf}
\end{eqnarray}

With $\beta f({\bf q})$ given by Eq.\ (\ref{Bf}) the integral in Eq.\ (\ref{Zn6}) can be evaluated using the saddle point method. For this
purpose, the minimum of the function $f\left( m^{(A)},m^{(B)},q^{(A)},q^{(B)}\right)$ should be found, and the necessary condition for
the existence of extremum leads to the following set of self-consistent equations for the order parameters,
\begin{eqnarray}
\frac{\partial f}{\partial m^{(A)}} =0 &\Leftrightarrow& m^{(A)}= \frac{1}{\sqrt{2\pi}} \int dz e^{-z^{2}/2} \sum_{i=1}^{N} v_{i}^{(A)} \tanh \eta_{i}(z),
\nonumber\\
\frac{\partial f}{\partial q^{(A)}} =0 &\Leftrightarrow& q^{(A)}= \frac{1}{\sqrt{2\pi}} \int dz e^{-z^{2}/2} \sum_{i=1}^{N} v_{i}^{(A)} \tanh^{2}\eta_{i}(z),
\label{system}
\end{eqnarray}
and two similar equations for $m^{(B)}$, $q^{(B)}$.

\section{Critical temperatures for the ferromagnetic and spin glass transitions}

\subsection{General equations for the critical temperature}

For small $m^{(A)}$, $m^{(B)}$, $q^{(A)}$, $q^{(B)}$, after expanding the logarithm and evaluating the moments in Eq.\ (\ref{Bf}) the free
energy can be written as
\begin{eqnarray}
&& \beta f\left( m^{(A)},m^{(B)},q^{(A)},q^{(B)}\right)= \nonumber\\
&& \frac{K^{(A)} {\bf T}_{1}^{(A)}}{2}m^{(A)2}+\frac{K^{(B)} {\bf T}_{1}^{(B)}}{2}m^{(B)2} - 
\frac{K^{(A)} {\bf T}_{2}^{(A)}}{4}q^{(A)2}-\frac{K^{(B)} {\bf T}_{2}^{(B)}}{4}q^{(B)2} \nonumber\\
&& - \frac{1}{2}N \sum_{i=1}^{N} \left( K^{(A)}  {\bf T}_{1}^{(A)2}v_{i}^{(A)} m^{(A)} 
+ K^{(B)} {\bf T}_{1}^{(B)2} v_{i}^{(B)} m^{(B)} \right)^{2} \nonumber\\
&& + \frac{1}{4}N \sum_{i=1}^{N} \left( K^{(A)} {\bf T}_{2}^{(A)2} v_{i}^{(A)} q^{(A)} 
+ K^{(B)}{\bf T}_{2}^{(B)2}  v_{i}^{(B)} q^{(B)} \right)^{2}.
\label{Bflinear}
\end{eqnarray}
Then the system of equations in Eq.\ (\ref{system})
leads to the following system of linear equations for $m^{(A)}$, $m^{(B)}$, $q^{(A)}$, $q^{(B)}$,
\begin{eqnarray}
\left( 1-NK^{(A)} {\bf T}_{1}^{(A)} \sum_{i=1}^{N} v_{i}^{(A)2} \right) m^{(A)} 
-NK^{(B)}{\bf T}_{1}^{(B)}\left( \sum_{i=1}^{N} v_{i}^{(A)}v_{i}^{(B)}\right) m^{(B)} &=&0 \nonumber\\
-NK^{(A)}{\bf T}_{1}^{(A)}\left( \sum_{i=1}^{N} v_{i}^{(A)}v_{i}^{(B)}\right) m^{(A)}
+\left( 1-NK^{(B)} {\bf T}_{1}^{(B)} \sum_{i=1}^{N} v_{i}^{(B)2} \right) m^{(B)} &=& 0.\nonumber\\
\left( 1-NK^{(A)} {\bf T}_{2}^{(A)} \sum_{i=1}^{N} v_{i}^{(A)2} \right) q^{(A)} 
-NK^{(B)}{\bf T}_{2}^{(B)}\left( \sum_{i=1}^{N} v_{i}^{(A)}v_{i}^{(B)}\right) q^{(B)} &=&0 \nonumber\\
-NK^{(A)}{\bf T}_{2}^{(A)}\left( \sum_{i=1}^{N} v_{i}^{(A)}v_{i}^{(B)}\right) q^{(A)}
+\left( 1-NK^{(B)} {\bf T}_{2}^{(B)} \sum_{i=1}^{N} v_{i}^{(B)2} \right) q^{(B)} &=& 0.\nonumber\\
&& \label{mAmB}
\end{eqnarray}
Non-zero solutions of the system of Eq.\ (\ref{mAmB}) exist if the determinant is zero. Due to the block structure of Eq.\ (\ref{mAmB})
this condition is equivalent to the requirement that the determinant of the system of the first two equations of the above system is zero
or the determinant of the last two equations of the above system is zero. From the former condition the critical temperature for the FM
transition can be evaluated: the corresponding equation for the critical temperature is quadratic with respect to $\tanh \beta J^{(A)}$, $\tanh \beta J^{(B)}$, thus it has two
solutions of which that with a higher value corresponds to $T_{c}^{FM}$. From the latter condition the critical temperature for the SG transition
can be evaluated: the corresponding equation for the critical temperature is biquadratic with respect to $\tanh \beta J^{(A)}$, $\tanh \beta J^{(B)}$, thus it has four
solutions of which the real solution with the highest value corresponds to $T_{c}^{SG}$. In Sec.\ 4.2 - 4.5 both critical temperatures are evaluated
for the MN with random ER and SF layers.

From the above-mentioned procedure, for fixed model parameters, it is possible to determine on the phase diagram the boundary between the paramagnetic
andt the FM or SG phase, depending on which of the temperatures $T_{c}^{FM}$, $T_{c}^{SG}$ is higher. The boundary between the FM and SG phases
runs along the Almeida-Thouless line at which the FM phase becomes unstable against the occurrence of the reentrant SG phase
as the temperature is lowered \cite{Almeida78}. Determination of the location of this line on the phase diagram is not straightforward in the case of
the Ising model on MNs for which, in contrast with SG models considered so far, the RS solution is characterized by two sets of order parameters 
$m^{(A)}$, $q^{(A)} \ldots$ and $m^{(B)}$, $q^{(B)}\ldots $ connected with the two layers $G^{(A)}$, $G^{(B)}$; moreover, it requires numerical solution of a set
of four nonlinear equations in Eq.\ (\ref{system}). Thus, this problem is left for future research and full phase diagrams for the model under study 
are not presented in this paper.

The space of parameters for the Ising model on a MN is large and comprises $J^{(A)}$, $J^{(B)}$ and the parameters of the layers,
e.g., in the case of SF layers, $K^{(A)}$, $K^{(B)}$, $\gamma^{(A)}$, $\gamma^{(B)}$, $r^{(A)}$, $r^{(B)}$ as well as the correlation 
between the degrees of nodes within the layers $k_{i}^{(A)}$, $k_{i}^{(B)}$, $i=1,2,\ldots N$. In order to constraint the number of 
independent parameters henceforth all formulae are obtained under the assumptions $J^{(A)}=J^{(B)}=J$, $r^{(A)}=1$ (i.e., in the layer $G^{(A)}$
the interactions are purely FM) and $r^{(B)}=0$ (i.e., in the layer $G^{(B)}$ all interactions are purely AFM); moreover, mostly the
case $K^{(A)}=K^{(B)}$ is considered. Under such assumptions it is possible to study the, probably most intriguing, problem of the influence of the 
difference between the distributions of the degerees of nodes $p_{k^{(A)}}$, $p_{k^{(B)}}$ and of the correlation between degrees of nodes within
different layers on the FM and SG transition in the case of balanced FM and AFM interactions. 

\subsection{Random Erd\"os-R\'enyi layers}

For random ER layers there is $v_{i}^{(A)}=v_{i}^{(B)}=1/N$, $i=1,2,\ldots N$. From
the definition of the magnetizations, Eq.\ (\ref{ordpar}), as well as from the first two equations in Eq.\ (\ref{mAmB}) follows that
$m^{(A)}=m^{(B)}$ and the equation for the FM critical temperature is linear rather than quadratic with respect to $\tanh\beta J$. The result is
\begin{equation}
T_{c}^{FM}= J{\rm atanh}^{-1} \left[ K^{(A)} -K^{(B)}\right]^{-1},
\label{TfmER}
\end{equation}
thus $T_{c}^{FM} <0$ for $K^{(A)}< K^{(B)}$ and if the density of edges corresponding to FM interactions is smaller than that of edges
corresponding to AFM interactions the transition to the FM phase cannot occur. From 
the definition of the SG order parameters, Eq.\ (\ref{ordpar1}), as well as from the last two equations in Eq.\ (\ref{mAmB})
follows that $q^{(A)}=q^{(B)}$ and the equation for the SG critical temperature is linear with respect to $\tanh^{2}\beta J$. The result is
\begin{equation}
T_{c}^{SG}= J{\rm atanh}^{-1} \left[ K^{(A)} + K^{(B)}\right]^{-1/2},
\label{TsgER}
\end{equation}
which is always finite and positive. Thus for $K^{(A)}< K^{(B)}$ there is only SG transition from the paramagnetic phase, and in the
opposite case the transition can be either to the SG or to the FM phase, depending on which of the critical temperatures, $T_{c}^{SG}$ or
$T_{c}^{FM}$ is higher for given $K^{(A)}$, $K^{(B)}$.

\subsection{Independent scale-free layers}

The MN with two independent SF layers is generated by randomly and independently assigning to the nodes $i=1,2,\ldots N$ the weights from 
a set $\left\{ v_{k}: v_{k}=k^{-\mu^{(A)}}/\zeta_{N}( \mu^{(A)}) \right\}$, $k=1,2,\ldots N$ to generate the layer $G^{(A)}$ and from
a set $\left\{ v_{l}: v_{l}=l^{-\mu^{(B)}}/\zeta_{N}( \mu^{(B)}) \right\}$, $l=1,2,\ldots N$ to generate the layer $G^{(B)}$. In this way,
the sequences of weights $v_{i}^{(A)}$, $v_{i}^{(B)}$ are uncorrelated. 
As a result, in Eq.\ (\ref{mAmB}) the sum over the products of weights can be approximated by its expected value,
$ N\sum_{i=1}^{N} v_{i}^{(A)}v_{i}^{(B)} \approx N \langle \sum_{i=1}^{N} v_{i}^{(A)}v_{i}^{(B)} \rangle = 1$
\cite{Krawiecki17}. This approximation is valid in typical cases of MNs with independently generated layers, and is applied instead of
averaging the partition function in Eq.\ (\ref{Zn}) over a class of MNs with mutually independent sequences of weights
assigned to nodes when generating different layers.
Besides, for $\mu^{(A)}<1/2$ ($\gamma^{(A)}>3$), $\mu^{(B)}<1/2$ ($\gamma^{(B)}>3$) by approximating the
sum with an integral it is obtained that
$ N\sum_{i=1}^{N}v_{i}^{(A)2}= N\sum_{k=1}^{N}v_{k}^{2}\approx \frac{\left(1-\mu^{(A)}\right)^2}{1-2\mu^{(A)}}
=\frac{\left( \gamma^{(A)}-2 \right)^{2}}{\left( \gamma^{(A)}-1\right) \left( \gamma^{(A)}-3\right)}$
and similarly for $\sum_{i=1}^{N} v_{i}^{(B)2}$.

The critical temperature for the FM transition can be evaluated from the condition that the determinant of a system of the first two equations in Eq.\ (\ref{mAmB})
is zero, which leads to
\begin{equation}
T_{c}^{FM}=J {\rm atanh}^{-1} \left\{
\frac{ K^{(A)}  \frac{\left(1-\mu^{(A)}\right)^2}{1-2\mu^{(A)}} - K^{(B)} \frac{\left(1-\mu^{(B)}\right)^2}{1-2\mu^{(B)}}-\sqrt{\Delta}}
{2 K^{(A)}K^{(B)} \left[ 1- \frac{\left(1-\mu^{(A)}\right)^2}{1-2\mu^{(A)}} \frac{\left(1-\mu^{(B)}\right)^2}{1-2\mu^{(B)}} \right]}
\right\},
\label{TcFM1}
\end{equation}
where
\begin{displaymath}
\Delta = \left[ K^{(A)}  \frac{\left(1-\mu^{(A)}\right)^2}{1-2\mu^{(A)}} + K^{(B)} \frac{\left(1-\mu^{(B)}\right)^2}{1-2\mu^{(B)}} \right]^2
-4 K^{(A)}K^{(B)}.
\end{displaymath}
In particular, for the two layers with identical distributions of the degrees of nodes $p_{k^{(A)}} = p_{k^{(B)}}=p_{k}$, i.e., with
$K^{(A)}=K^{(B)}=K$, $\mu^{(A)}=\mu^{(B)}=\mu$,
\begin{equation}
T_{c}^{FM}=J {\rm atanh}^{-1}\left[ K^{-1}\left( \frac{\left(1-\mu \right)^4}{\left( 1-2\mu\right)^{2}} -1\right)^{-1/2} \right],
\label{TcFM1eq}
\end{equation}
i.e., in this case $T_{c}^{FM}>0$ for $0< \mu <1/2$ ($\gamma >3$) and $T_{c}^{FM}\rightarrow \infty$ for $\mu \rightarrow 1/2$ ($\gamma \rightarrow 3$).

The temperature for the SG transition can be evaluated from the condition that the determinant of a system of the last two equations in Eq.\ (\ref{mAmB})
is zero, which leads to
\begin{equation}
T_{c}^{SG}=J {\rm atanh}^{-1} \left\{
\frac{ K^{(A)}  \frac{\left(1-\mu^{(A)}\right)^2}{1-2\mu^{(A)}} + K^{(B)} \frac{\left(1-\mu^{(B)}\right)^2}{1-2\mu^{(B)}}-\sqrt{\Delta}}
{2 K^{(A)}K^{(B)} \left[ \frac{\left(1-\mu^{(A)}\right)^2}{1-2\mu^{(A)}} \frac{\left(1-\mu^{(B)}\right)^2}{1-2\mu^{(B)}} -1 \right]}
\right\}^{1/2},
\label{TcSG1}
\end{equation}
where
\begin{displaymath}
\Delta = \left[ K^{(A)}  \frac{\left(1-\mu^{(A)}\right)^2}{1-2\mu^{(A)}} - K^{(B)} \frac{\left(1-\mu^{(B)}\right)^2}{1-2\mu^{(B)}} \right]^2
+4 K^{(A)}K^{(B)}.
\end{displaymath}
In particular, for the two layers with identical distributions of the degrees of nodes
\begin{equation}
T_{c}^{SG}=J {\rm atanh}^{-1}\left[ K^{-1/2}\left( \frac{\left(1-\mu \right)^2}{1-2\mu} +1\right)^{-1/2} \right],
\label{TcSG1eq}
\end{equation}
i.e., in this case again $T_{c}^{SG}>0$ for $0< \mu <1/2$ ($\gamma >3$) and $T_{c}^{SG}\rightarrow \infty$ for $\mu \rightarrow 1/2$ 
($\gamma \rightarrow 3$).

\begin{figure}[bt]
\centerline{\includegraphics[width=1.0\textwidth]{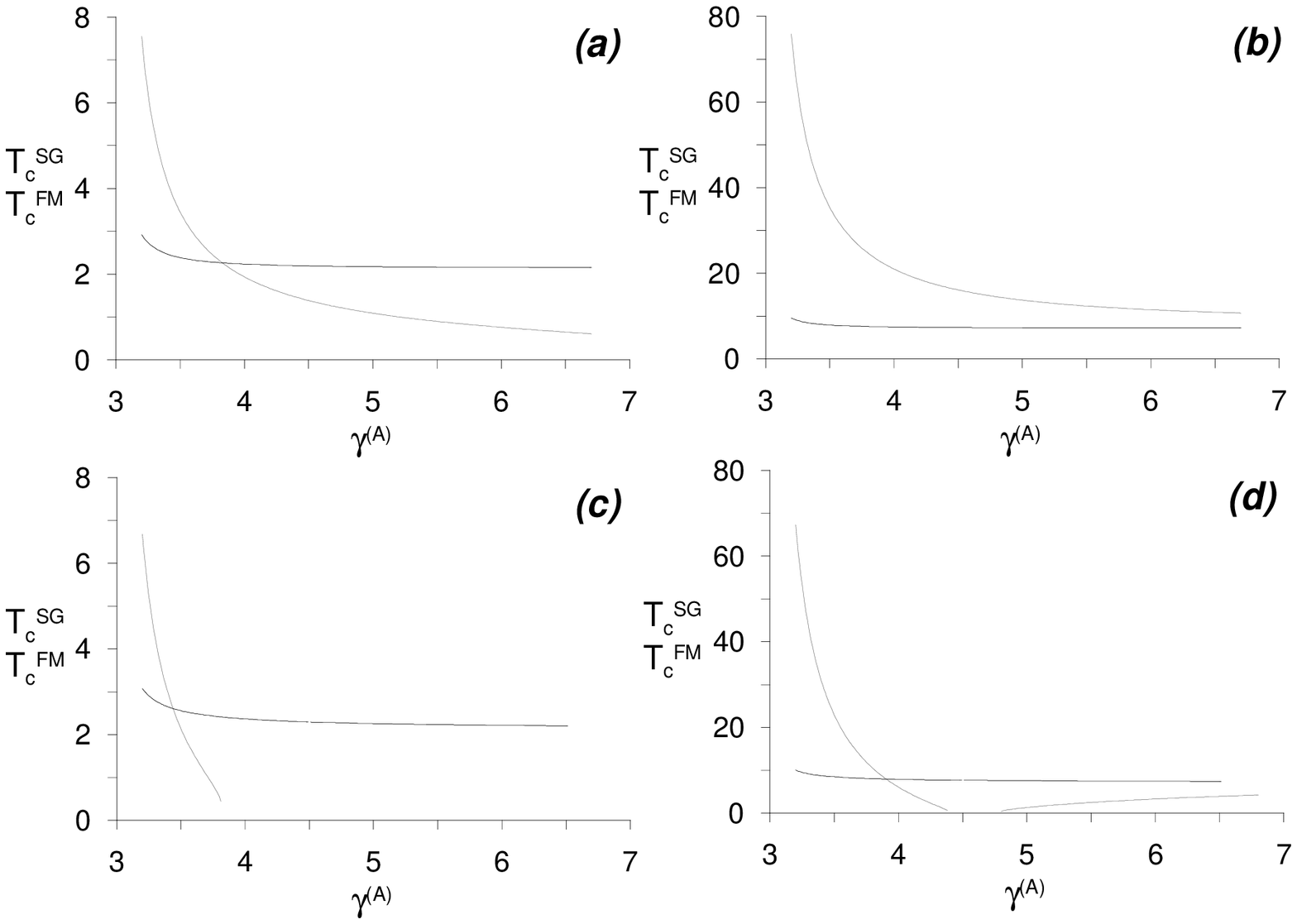}}
\caption{The critical temperatures for the SG transition $T_{c}^{SG}$ (black solid lines) and for the FM transition $T_{c}^{FM}$ (gray solid lines) 
vs.\ $\gamma^{(A)}$, for $\gamma^{(B)}=4.5$ and for (a,b) the MN with independent SF layers with (a) $K^{(A)}=K^{(B)} =2.5$, (b)
$K^{(A)}=K^{(B)}=25$, results of Eq.\ (\ref{TcSG1}), Eq.\ (\ref{TcFM1}), respectively, and for (c,d) maximally correlated SF layers
with (c) $K^{(A)}=K^{(B)} =2.5$, (d) $K^{(A)}=K^{(B)}=25$, results of Eq.\ (\ref{TcSG2}), Eq.\ (\ref{TcFM2}), respectively}
\end{figure}

As an example, in Fig.\ 1(a,b) the critical temperatures $T_{c}^{FM}$, $T_{c}^{SG}$ evaluated from Eq.\ (\ref{TcFM1}) and Eq.\ (\ref{TcSG1}),
respectively, are shown for fixed $\gamma^{(A)}=4.5$. It is assumed that $K^{(A)}=K^{(B)}$ which means that the FM and AFM
interactions in the MN are balanced (the number of edges corresponding to $J>0$ in the layer $G^{(A)}$ is equal to that corresponding to $-J<0$
in the layer $G^{(B)}$); Fig.\ 1(a) is for small and Fig.\ 1(b) for high mean degree of nodes within both layers. In the Ising model on SF networks
only SG transition occurs in the case of balanced FM and AFM interactions \cite{Kim05}. Here, in contrast, depending on $\gamma^{(A)}$
and the mean degree of nodes transition to the SG or FM phase can occur from the paramagnetic phase. The critical temperature remains finite 
for $\gamma^{(A)}>3$. As $\gamma^{(A)}\rightarrow 3$ (from above),
i.e., as the second moment $\langle k^{(A)}\rangle^{2}$ of the distribution $p_{k^{(A)}}$ diverges, $T_{c}^{FM}\rightarrow \infty$ and 
the transition is to the FM phase. This is probably due to presence of a large number of hubs (nodes with high degree) within the layer $G^{(A)}$
interacting via FM interactions with many neighbors and thus enforcing global order. For larger $\gamma^{(A)}$ and small $K^{(A)}=K^{(B)}$
the transition is to the SG phase, and a tricritical point occurs on the phase diagram (Fig.\ 1(a)). In contrast, for large $K^{(A)}=K^{(B)}$ there is
$T_{c}^{FM}>T_{c}^{SG}$ in the whole range of $\gamma^{(A)}$ and the transition is always to the FM phase. In particular, even if the 
distributions of the degrees of nodes within both layers are the same, $p_{k^{(A)}}=p_{k^{(B)}}$ and thus $\gamma^{(A)}=\gamma^{(B)}$,
for small $K^{(A)}=K^{(B)}$ the transition is to the SG phase and for high $K^{(A)}=K^{(B)}$ to the FM phase, which is in marked contrast with
the above-mentioned case of the Ising model on SF networks. The difference is due to the fact that in these two cases the FM and AFM 
interactions are not balanced in the same way. In the Insing model on SF networks, on average, half of edges attached to each node corresponds to 
$+J$ and half to $-J$ while in the model on MNs with independent layers there are many nodes with high degree within one layer (hubs) with many
attached edges corresponding, e.g., to FM interactions and with small degree within the other layer with only few attached edges corrseponding to
AFM interactions; i.e., in the latter case the FM and AFM interactions are balanced only globally but not locally. 
This result, confirmed via MC simulations in Sec.\ 4.6, emphasises the difference of the critical behavior 
between the Ising model on, possibly heterogeneous, networks and on MNs with separately (in particular, independently) generated layers. 

\subsection{Maximally correlated scale-free layers}

The MN with two maximally correlated SF layers is generated by randomly assigning to the nodes $i=1,2,\ldots N$ the weights 
$v_{i}^{(A)}=i^{-\mu^{(A)}}/\zeta_{N}( \mu^{(A)})$ to generate the layer $G^{(A)}$ and 
$v_{i}^{(B)}=i^{-\mu^{(B)}}/\zeta_{N}( \mu^{(B)})$ to generate the layer $G^{(B)}$. In this way, in the statistical ensemble of
MNs generated in this way the mean degrees of the consecutive nodes within each layer $\langle k_{i}^{(A)} \rangle \propto v_{i}^{(A)}$, 
$\langle k_{i}^{(B)} \rangle \propto v_{i}^{(B)}$ \cite{Lee04} are maximally correlated; 
As a result, the nodes which have high degree within one layer have also, on average, high degree in the other layer and
vice versa. Then, for $\mu^{(A)}<1/2$ ($\gamma^{(A)}>3$), $\mu^{(B)}<1/2$ ($\gamma^{(B)}>3$) approximating the
sum with an integral there is
$ N\sum_{i=1}^{N} v_{i}^{(A)}v_{i}^{(B)} \approx \frac{\left(1-\mu^{(A)}\right) \left(1-\mu^{(B)}\right)}{1-\left( \mu^{(A)}+ \mu^{(B)} \right)}$
in Eq.\ (\ref{mAmB}). Since all MNs generated in this way are equivalent up to the permutation of the indices of nodes,
no further averaging as in the case of independent SF layers is necessary.

The temperature for the FM transition can be evaluated from the condition that the determinant of a system of the first two equations in Eq.\ (\ref{mAmB})
is zero, which leads to
\begin{equation}
T_{c}^{FM}=J {\rm atanh}^{-1} \left\{
\frac{ K^{(A)}  \frac{\left(1-\mu^{(A)}\right)^2}{1-2\mu^{(A)}} - K^{(B)} \frac{\left(1-\mu^{(B)}\right)^2}{1-2\mu^{(B)}}-\sqrt{\Delta}}
{2 K^{(A)}K^{(B)} \left[ 
\frac{\left(1-\mu^{(A)}\right)^2 \left(1-\mu^{(B)}\right)^2}{\left[ 1-\left( \mu^{(A)} + \mu^{(B)} \right)\right]^{2}}
- \frac{\left(1-\mu^{(A)}\right)^2}{1-2\mu^{(A)}} \frac{\left(1-\mu^{(B)}\right)^2}{1-2\mu^{(B)}} \right]}
\right\},
\label{TcFM2}
\end{equation}
where
\begin{displaymath}
\Delta = \left[ K^{(A)}  \frac{\left(1-\mu^{(A)}\right)^2}{1-2\mu^{(A)}} + K^{(B)} \frac{\left(1-\mu^{(B)}\right)^2}{1-2\mu^{(B)}} \right]^2
-4 K^{(A)}K^{(B)} \frac{\left(1-\mu^{(A)}\right)^2 \left(1-\mu^{(B)}\right)^2}{\left[ 1-\left( \mu^{(A)} + \mu^{(B)} \right)\right]^{2}}.
\end{displaymath}
In particular, it can be seen that for the two layers with identical distributions of the degrees of nodes the above-mentioned determinant is
equal to 1 for any $\beta$, thus transition to the FM state cannot occur.

The temperature for the SG transition can be evaluated from the condition that the determinant of a system of the last two equations in Eq.\ (\ref{mAmB})
is zero, which leads to
\begin{equation}
T_{c}^{SG}=J {\rm atanh}^{-1} \left\{
\frac{ K^{(A)}  \frac{\left(1-\mu^{(A)}\right)^2}{1-2\mu^{(A)}} + K^{(B)} \frac{\left(1-\mu^{(B)}\right)^2}{1-2\mu^{(B)}}-\sqrt{\Delta}}
{2 K^{(A)}K^{(B)} \left[ \frac{\left(1-\mu^{(A)}\right)^2}{1-2\mu^{(A)}} \frac{\left(1-\mu^{(B)}\right)^2}{1-2\mu^{(B)}} 
-\frac{\left(1-\mu^{(A)}\right)^2 \left(1-\mu^{(B)}\right)^2}{\left[ 1-\left( \mu^{(A)} + \mu^{(B)} \right)\right]^{2}} \right]}
\right\}^{1/2},
\label{TcSG2}
\end{equation}
where
\begin{displaymath}
\Delta = \left[ K^{(A)}  \frac{\left(1-\mu^{(A)}\right)^2}{1-2\mu^{(A)}} - K^{(B)} \frac{\left(1-\mu^{(B)}\right)^2}{1-2\mu^{(B)}} \right]^2
+4 K^{(A)}K^{(B)} \frac{\left(1-\mu^{(A)}\right)^2 \left(1-\mu^{(B)}\right)^2}{\left[ 1-\left( \mu^{(A)} + \mu^{(B)} \right)\right]^{2}}.
\end{displaymath}
In particular, for the two layers with identical distributions of the degrees of nodes
\begin{equation}
T_{c}^{SG}=J {\rm atanh}^{-1} \left( \sqrt{\frac{1-2\mu}{2K}}\frac{1}{1-\mu} \right),
\label{TcSG2eq}
\end{equation}
i.e., in this case again $T_{c}^{SG}>0$ for $0< \mu <1/2$ ($\gamma >3$) and $T_{c}^{SG}\rightarrow \infty$ for $\mu \rightarrow 1/2$ 
($\gamma \rightarrow 3$).

As an example, in Fig.\ 1(c,d) the critical temperatures $T_{c}^{FM}$, $T_{c}^{SG}$ evaluated from Eq.\ (\ref{TcFM2}) and Eq.\ (\ref{TcSG2}),
respectively, are presented for fixed $\gamma^{(A)}=4.5$ and small (Fig.\ 1(c)) and high (Fig.\ 1(d)) value of $K^{(A)}=K^{(B)}$.
As in the case of MN with independent layers for $\gamma^{(A)}\rightarrow 3$ there is $T_{c}^{FM}\rightarrow \infty$ and the transition is to the FM phase.
However, for larger $\gamma^{(A)}$ the transition is to the SG phase, independently of the mean degrees of nodes within layers; thus, there is
always the tricritical point on the phase diagram. In particular, in the case of MN with identical distributions of the degrees of nodes within both layers 
the transition is always to the SG phase, as in the case of the Ising model on SF networks with balanced FM and AFM interactions.
This similarity is due to the fact that in the case of the Ising model on a MN with maximally correlated layers the same nodes have high (or small)
degree within both layers, thus the FM and AFM interactions are balanced both globally and locally.

\subsection{Minimally correlated scale-free layers}

The MN with two minimally correlated SF layers is generated by randomly assigning to the nodes $i=1,2,\ldots N$ the weights 
$v_{i}=i^{-\mu^{(A)}}/\zeta_{N}(\mu^{(A)})$, $v_{N-i}^{(B)}=i^{-\mu^{(B)}}/\zeta_{N}(\mu^{(B)})$. As a result, 
 the nodes which have high degree within one layer have, on average, low degree in the other layer and
$N\sum_{i=1}^{N} v_{i}^{(A)}v_{i}^{(B)} \stackrel{N\rightarrow \infty}{\rightarrow} 0$ in Eq.\ (\ref{mAmB}).
Hence, the MN is effectively decomposed into two apparently non-interacting networks corresponding to the layers $G^{(A)}$, $G^{(B)}$, with
FM and AFM exchange interactions, respectively. As a result,
the system of equations in Eq.\ (\ref{mAmB}) is decomposed into four independent equations. From the first two equations two 
critical temperatures for the FM transition can be obtained as
\begin{eqnarray}
&& T_{c}^{(A)FM}= J {\rm atanh}^{-1}\left[ \frac{1}{K^{(A)}}\frac{1-2\mu^{(A)}}{\left( 1-\mu^{(A)}\right)^{2}} \right], \nonumber\\
&& T_{c}^{(B)FM}= - J {\rm atanh}^{-1}\left[ \frac{1}{K^{(B)}}\frac{1-2\mu^{(B)}}{\left( 1-\mu^{(B)}\right)^{2}} \right],
\end{eqnarray}
of which $T_{c}^{(A)FM}$ is positive and finite for $1/2 < \mu^{(A)} <1$ while
$T_{c}^{(B)FM}<0$. From the last two equations in Eq.\ (\ref{mAmB}) the two critical temperatures
for the SG transition can be obtained as
\begin{eqnarray}
&& T_{c}^{(A)SG}= J {\rm atanh}^{-1}\left[ \sqrt{\frac{1-2\mu^{(A)}}{K^{(A)}}}\frac{1}{1-\mu^{(A)}} \right], \nonumber\\
&& T_{c}^{(B)SG}= J {\rm atanh}^{-1}\left[ \sqrt{\frac{1-2\mu^{(B)}}{K^{(B)}}}\frac{1}{1-\mu^{(B)}} \right],
\end{eqnarray}
which are finite and positive for $1/2 < \mu^{(A)} <1$, $1/2 < \mu^{(B)} <1$. There is always $T_{c}^{(A)SG} < T_{c}^{(A)FM}$,
thus, transition from the paramagnetic to the FM or SG phase
can occur at $T_{c}= \max \left\{ T_{c}^{(A)FM}, T_{c}^{(B)SG}\right\}$ as the temperature is lowered, depending on the parameters 
$K^{(A)}$, $K^{(B)}$, $\mu^{(A)}$, $\mu^{(B)}$.

\subsection{Comparison with Monte Carlo simulations}

\begin{figure}
\centerline{\includegraphics[width=1.0\textwidth]{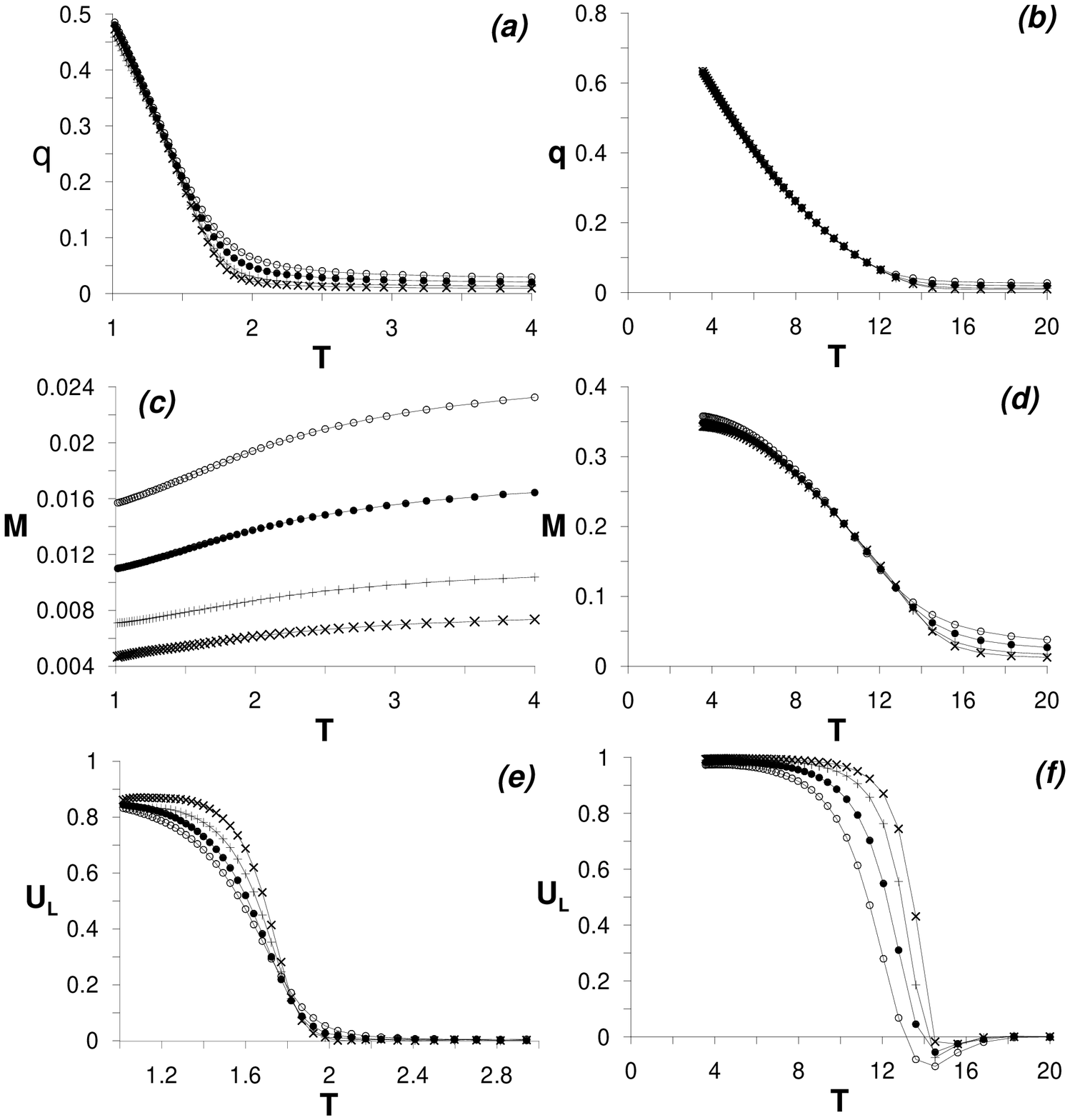}}
\caption{Results of MC simulations for the MN with independent SF layers with $\gamma^{(A)}=\gamma^{(B)}=4.5$, (a,c,e) $K^{(A)}=K^{(B)}=2$, 
(b,d,f) $K^{(A)}=K^{(B)}=25$: (a,b) the overlap order parameter $q$, Eq.\ (\ref{qoverlap}), vs.\ $T$, (c,d) the magnetization $M$ vs.\ $T$,
(e,f) the Binder cumulant $U_{L}$, Eq.\ (\ref{B}), vs.\ $T$, for ($\circ$) $N=1000$, ($\cdot$) $N=2000$, ($+$) $N=5000$, ($\times$) $N=10000$.}
\end{figure}

In order to verify, at least partly, the theoretical predictions of Sec.\ 4.3 MC simulations of the Ising model on MNs with independent SF layers were performed
using the Metropolis algorithm and the parallel tempering (replica exchange) method \cite{Hukushima96,Katzgraber01,Katzgraber02} 
in the form described in Ref.\ \cite{Bartolozzi06}.
As the order parameter for the SG transition the absolute value $\left| q\right|$ of the overlap parameter
\begin{equation}
q=\left[ \langle \frac{1}{N}\sum_{i=1}^{N} s_{i}^{\alpha} s_{i}^{\beta} \rangle_{t} \right]_{av} \equiv \left[ \langle \tilde{q} \rangle_{t} \right]_{av} 
\label{qoverlap}
\end{equation}
is used, where $\alpha$, $\beta$ denote two copies (replicas) of the system simulated independently with random initial conditions,
$\langle \cdot \rangle_{t}$ denotes the time average for the system on a given MN and $\left[ \cdot \right]_{av}$, as in Sec.\ 3, denotes average over
different realizations of the MN with independent SF layers. The critical temperature for the SG transition $T_{c}^{SG}$
can be determined from the intersection point of the Binder cumulants $U_{L}$ vs.\ $T$ for different $N$ \cite{Binder97}, where
\begin{equation}
U_{L}=\frac{1}{2}\left[ 3-\frac{\langle \tilde{q}^{4} \rangle_{t}}{\langle \tilde{q}^{2}\rangle_{t}^{2}} \right]_{av}.
\label{B}
\end{equation}
Below the critical
temperature for the transition from the paramagnetic to the SG phase $\left| q\right|$ increases from zero. In contrast, the 
absolute value $\left| M\right|$ of the magnetization $M=\left[ \langle N^{-1}\sum_{i=1}^{N} s_{i}\rangle_{t}\right]_{av}$,
which is the order parameter for the FM transition, should remain close to zero. 
On the other hand, below the critical temperature for the transition from the paramagnetic to the FM phase both $\left| q\right|$ and $\left| M \right|$ should
increase from zero.

An important prediction in Sec.\ 4.3 is that in the case of MN with independent SF layers with identical distributions of the degrees of nodes
$p_{k^{(A)}} =p_{k^{(B)}}$ for a certain range
of $\gamma^{(A)}=\gamma^{(B)}$ the transition from the paramegnetic to the SG phase occurs only for small $K^{(A)}=K^{(B)}$ while for
layers with higher mean degree of nodes transition to the FM state is expected. 
This scenario is in fact observed in MC simulations; results for $\gamma^{(A)}=\gamma^{(B)}=4.5$ and for two values of
$K^{(A)}=K^{(B)}$ are shown in Fig.\ 2. For $K^{(A)}=K^{(B)}=2$ the SG transition occurs: for $T< T_{c}^{SG}$ increase of $\left| q\right|$ can
be seen (Fig.\ 2(a)) while $\left| M\right|$ remains small and even decreases (Fig.\ 2(c)). Also the value of $T_{c}^{SG} =1.85 \pm 0.1$ estimated from the MC
simulations (Fig.\ 2(e)) is close to $T_{c}^{SG}=1.92$ evaluated from Eq.\ (\ref{TcSG1eq}). In contrast, for  $K^{(A)}=K^{(B)}=25$ 
the FM transition occurs: both $\left| q\right|$ and $\left|M\right|$ increase for $T< T_{c}^{FM}$ (Fig.\ 2(b,d)) and the value of the FM critical temperature 
obtained from the MC simulations can be assessed as $T_{c}^{FM} = 15 \pm 2$ from Fig.\ 2(d), which is again with agreement with
$T_{c}^{FM}=16.13$ evaluated from Eq.\ (\ref{TcFM1eq}). Thus, in this particular case theoretical predictions based on the RS solution are confirmed by MC simulations.

\section{Critical exponents for the spin glass transition}

Below the transition point from the paramagnetic to the SG phase the SG order parameter increases from zero while the magnetization 
remains close to zero. Bleow $T_{c}^{SG}$ the SG order parameter is expected to scale as $\varepsilon^{\beta}$, 
where $\varepsilon =\left( T_{c}^{SG}-T\right)/T_{c}^{SG}$. In Ref.\ \cite{Kim05} it was shown that in the case of the Ising model on 
SF networks the scaling exponent $\beta$ can depend on the parameters of the distribution of the degrees of nodes. In this section
this exponent is determined for the Ising model on MNs with independent layers using a semi-analytic procedure.

Let us start with the Ising model on a MN with two random ER layers.
In the case of the transition from the paramagnetic to the SG phase $m^{(A)}=m^{(B)}=0$ below the critical temperature. Taking into 
account that $v_{i}^{(A)}=v_{i}^{(B)}=1/N$, $i=1,2,\ldots N$, ${\bf T}_{2}^{(A)}={\bf T}_{2}^{(B)}= {\bf T}_{2} =\tanh^{2}\beta  J$ from Eq.\ (\ref{T1T2})
and $q^{(A)}=q^{(B)}=q$ (see Sec.\ 4.2) the equations
for the SG order parameter, the last two equations in Eq.\ (\ref{system}), are reduced to a single equation
\begin{equation}
q= \frac{1}{\sqrt{2\pi}} \int dz e^{-z^{2}/2} \tanh^{2}\left[ z\sqrt{\left( K^{(A)}+K^{(B)} \right) {\bf T}_{2}q } \right].
\end{equation}
In the vicinity of $T_{c}^{SG}$ it is possible to expand the $\tanh \left( \cdot\right)$ function with respect to $q$. After evaluating the momenta and
retaining only the lowest-order nonlinear term it is obtained that
\begin{equation}
\left( \frac{1}{\left( K^{(A)}+K^{(B)} \right) {\bf T}_{2}} - \frac{1}{\left( K^{(A)}+K^{(B)} \right) {\bf T}_{2,c}^{SG}} \right) Q =-2Q^{2},
\end{equation}
where 
\begin{displaymath}
1= \left( K^{(A)}+K^{(B)} \right) {\bf T}_{2,c}^{SG} =  \left( K^{(A)}+K^{(B)} \right) \tanh^{2} \frac{J}{T_{c}^{SG}}
\end{displaymath}
(see Eq.\ (\ref{TsgER})) and $Q=  \left( K^{(A)}+K^{(B)} \right) {\bf T}_{2}$. Hence, $q\propto \varepsilon$ just below the SG transition temperature,
as in the case of the SG transition in the Ising model on random ER graphs
\cite{Viana85}.

Let us now consider the Ising model on a MN with two independent SF layers.
The SG transition temperature is finite, and thus the scaling behavior of the order parameters below $T_{c}^{SG}$ can be determined,
for $\gamma^{(A)}>3$, $\gamma^{(B)}>3$.
In the case of the transition from the paramagnetic to the SG phase $m^{(A)}=m^{(B)}=0$ below the critical temperature and the equations
for the SG order parameter, the last two equations in Eq.\ (\ref{system}), are
\begin{equation}
q^{(A)}= \frac{1}{\sqrt{2\pi}} \int dz e^{-z^{2}/2} \sum_{i=1}^{N} v_{i}^{(A)} 
\tanh^{2}\left[  z \sqrt{  N  \left( K^{(A)}{\bf T}_{2}^{(A)} v_{i}^{(A)} q^{(A)} + K^{(B)} {\bf T}_{2}^{(B)} v_{i}^{(B)} q^{(B)}\right) } \right],
\label{qAsg1}
\end{equation}
and analogous equation for $q^{(B)}$. Unfortunately, in this case it is not possible simply to expand the $\tanh \left( \cdot\right)$ function with respect to $q$
due to the occurrence of the terms like $N^{-1}\sum_{i=1}^{N} v_{i}^{(A)3}$, etc., under the integral which diverge even if the second moments of
the distribution of the weights associated with each layer are finite. Nevertheless, as shown in Appendix B
the sum over the indices of nodes on the right-hand side of Eq.\ (\ref{qAsg1}) can be represented in a form of a converging series expansion
with respect to $q^{(A)}$, $q^{(B)}$.

First, let us note that in the case of independent SF layers the sum over the indices of nodes in Eq.\ (\ref{qAsg1}) can be
replaced by its expected value, similarily as in Sec.\ 4.3, and then approximated by an integral,
\begin{eqnarray}
&& \sum_{i=1}^{N} v_{i}^{(A)} 
\tanh^{2}\left[  z \sqrt{  N  \left( K^{(A)}{\bf T}_{2}^{(A)} v_{i}^{(A)} q^{(A)} + K^{(B)} {\bf T}_{2}^{(B)} v_{i}^{(B)} q^{(B)}\right) } \right] 
\approx \nonumber\\
&& \sum_{i=1}^{N}  N^{-2} \sum_{k=1}^{N}\sum_{l=1}^{N} v_{k}
\tanh^{2}\left[  z \sqrt{  N  \left( K^{(A)}{\bf T}_{2}^{(A)} v_{k} q^{(A)} + K^{(B)} {\bf T}_{2}^{(B)} v_{l} q^{(B)}\right) } \right]
\approx \nonumber\\
&& \frac{1-\mu^{(A)}}{N^{2}} \int_{1}^{N}\int_{1}^{N} dy_{k} dy_{l} \left( \frac{N}{y_{k}}\right)^{\mu^{(A)}}
\tanh^{2}\left[  z \sqrt{  \left( \frac{N}{y_{k}}\right)^{\mu^{(A)}} Q^{(A)} +  \left( \frac{N}{y_{l}}\right)^{\mu^{(B)}} Q^{(B)} } \right],
\nonumber
\end{eqnarray}
where $Q^{(A)}=\left( 1-\mu^{(A)}\right) K^{(A)}{\bf T}_{2}^{(A)} q^{(A)}$,  $Q^{(B)}=\left( 1-\mu^{(B)}\right) K^{(B)}{\bf T}_{2}^{(B)} q^{(B)}$.
In the limit $N\rightarrow \infty$ and after replacing the variables $u_{1}=z^{2}Q^{(A)} \left(N/y_{k}\right)^{\mu^{(A)}}$,
$u_{2}=z^{2}Q^{(B)} \left(N/y_{l}\right)^{\mu^{(B)}}$ Eq.\ (\ref{qAsg1}) becomes
\begin{eqnarray}
\frac{q^{(A)}}{1-\mu^{(A)}}& =& \frac{1}{\sqrt{2\pi}} \left( \gamma^{(A)}-1\right) \left( \gamma^{(B)}-1\right) 
\left( Q^{(A)}\right)^{\gamma^{(A)}-2} \left( Q^{(B)}\right)^{\gamma^{(B)}-1}
\times \nonumber\\
&&\int dz e^{-z^{2}/2} z^{2\left( \gamma^{(A)}+ \gamma^{(B)} -3\right)}
\int_{ z^{2}Q^{(A)}}^{\infty} \int_{z^{2}Q^{(B)}}^{\infty} 
\frac{u_{1}\tanh^{2} \sqrt{u_{1}+u_{2}} }{u_{1}^{\gamma^{(A)}} u_{2}^{\gamma^{(B)}}} du_{2} du_{1}.
\nonumber\\
&& \label{qAsg2}
\end{eqnarray}

The two-dimesional integral in Eq.\ (\ref{qAsg2}) can be evaluated using Eq.\ (\ref{s}) in Appendix A with 
$F\left( x_{1},x_{2}\right) = x_{1}\tanh^{2} \sqrt{x_{1}+x_{2}}$, and the result is given by Eq.\ (\ref{Sfinal}). Inserting this result in
Eq.\ (\ref{qAsg2}) and evaluating the moments it is obtained that
\begin{eqnarray}
\frac{Q^{(A)}}{K^{(A)}{\bf T}_{2}^{(A)}} &=& \frac{\left( \gamma^{(A)}-2\right)^{2}}{\left( \gamma^{(A)}-1\right)}\left( \gamma^{(B)}-1\right)
\times \nonumber\\
&&\left[
\sum_{n_{1}=0}^{\infty} \sum_{n_{2}=0}^{\infty} 
\frac{f_{n_{1},n_{2}}\left( 2\left(n_{1}+n_{2}\right)-3\right)!!}
{\left( n_{1}-\gamma^{(A)}+1\right) \left( n_{2}-\gamma^{(B)}+1\right)}\left( Q^{(A)}\right)^{n_{1}-1} \left( Q^{(B)}\right)^{n_{2}} 
\right. \nonumber\\
&-& \sum_{n_{1}=0}^{\infty} 
\frac{2^{\gamma^{(B)}+n_{1}-2}}{ n_{1}-\gamma^{(A)} +1} \frac{\Gamma\left( \gamma^{(B)}+n_{1}-\frac{3}{2}\right)}{\sqrt{\pi}}
I_{2}\left( \gamma^{(B)},n_{1}\right) \left(Q^{(A)}\right)^{n_{1}-1} \left( Q^{(B)}\right)^{\gamma^{(B)}-1} \nonumber\\
&-& \sum_{n_{2}=0}^{\infty} 
\frac{2^{\gamma^{(A)}+n_{2}-2}}{ n_{2}-\gamma^{(B)} +1} \frac{\Gamma\left( \gamma^{(A)}+n_{2}-\frac{3}{2}\right)}{\sqrt{\pi}}
I_{1}\left( \gamma^{(A)},n_{2}\right) \left( Q^{(A)}\right)^{\gamma^{(A)}-2} \left( Q^{(B)}\right)^{n_{2}} \nonumber\\
 &+& \left.2^{\gamma^{(A)}+\gamma^{(B)}-3}\frac{\Gamma\left( \gamma^{(A)}+\gamma^{(B)}-\frac{5}{2}\right)}{\sqrt{\pi}}
I\left( \gamma^{(A)},\gamma^{(B)}\right) \left( Q^{(A)}\right)^{ \gamma^{(A)}-2} \left( Q^{(B)}\right)^{\gamma^{(B)}-1}
\right],\nonumber\\
&& \label{qAsg3}
\end{eqnarray}
where $f_{n_{1},n_{2}}$ are given by Eq.\ (\ref{Taylorcoeff}), $I_{1}\left( \gamma^{(A)},n_{2}\right)$ by Eq.\ (\ref{I11}),
$I_{2}\left( \gamma^{(B)},n_{1}\right)$ by Eq.\ (\ref{I2}) and $I\left( \gamma^{(A)}, \gamma^{(B)}\right)$ by Eq.\ (\ref{I}) in Appendix B.
The complementary equation for $Q^{(B)}/K^{(B)}{\bf T}_{2}^{(B)}$ can be obtained from Eq.\ (\ref{qAsg3}) by replacing $(A)$ with $(B)$
and vice versa. 

\begin{figure}[bt]
\centerline{\includegraphics[width=0.6\textwidth]{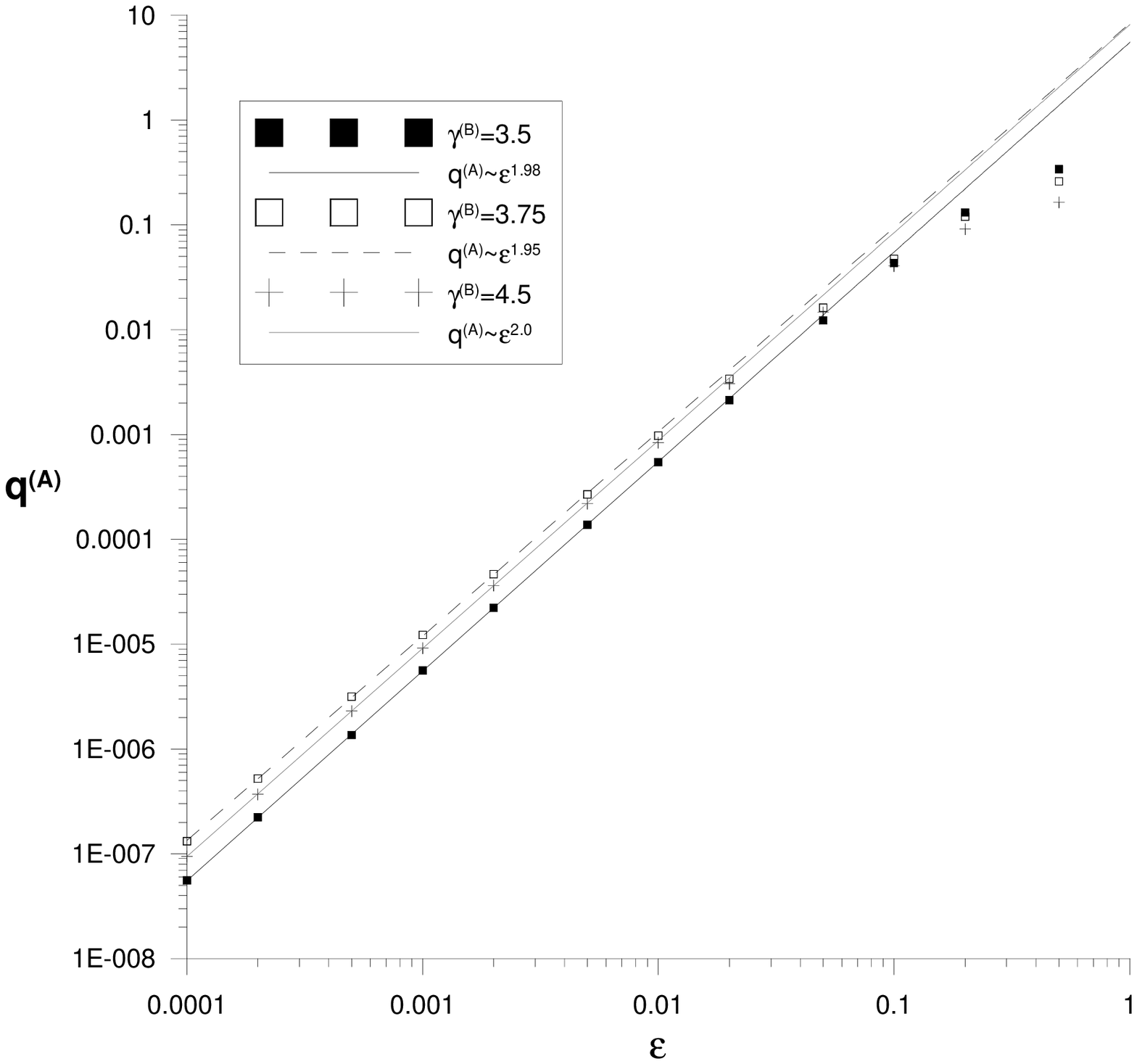}}
\caption{Symbols: $q^{(A)}$ vs.\ $\varepsilon = \left(T_{c}^{SG}-T\right)/T_{c}^{SG}$ 
in the neighborhood of the critical temperature for the SG transition
for the MN with independent SF layers with $\gamma^{(A)}=3.5$ and various $\gamma^{(B)}$ (see legend), the results were obtained from
numerical solution of Eq.\ (\ref{qAqBsg1}) for $\gamma^{(B)}<4$ and Eq.\ (\ref{qAqBsg2})
for $\gamma^{(B)}>4$ with $K^{(A)}=K^{(B)}=1$ (in all cases $T_{c}^{SG}>T_{c}^{FM}$). 
Lines: asymptotic (for $\varepsilon\rightarrow 0$) least-squares fits to the above-mentioned solutions
(see legend) in the form of the power scaling law $q^{(A)} \propto \varepsilon^{\beta}$.}
\end{figure}

Let us first consider the case when the SF layers have identical distributions of the degrees of nodes
$p_{k^{(A)}}=p_{k^{(B)}}$, with $K^{(A)}=K^{(B)}=K$, 
$\gamma^{(A)}=\gamma^{(B)}=\gamma$ ($\mu^{(A)}=\mu^{(B)}=\mu$).
Then the system of Eq.\ (\ref{qAsg1}) and thus also Eq.\ (\ref{qAsg3}) have solution with $q^{(A)}=q^{(B)}=q$.
In order to determine the critical behavior of $q$ near $T_{c}^{SG}$ it is necessary to retain only the lowest-order nonlinear term
on the right-hand side of Eq.\ (\ref{qAsg3}), thus the cases $3< \gamma <4$ and $\gamma>4$ should be considered separately.
For $3 <\gamma <4$ the following equation for $q$ is obtained from Eq.\ (\ref{qAsg3}),
\begin{equation}
\left( \frac{1}{K{\bf T}_{2}}-\frac{1}{K {\bf T}_{2,c}^{SG}} \right) Q = \frac{\left( \gamma-2\right)^{2}}{\left( \gamma-1\right)}2^{\gamma-1}
\frac{\Gamma\left( \gamma-\frac{3}{2}\right)}{\sqrt{\pi}} D\left(\gamma\right) Q^{\gamma-2},
\label{qAqBsg1eq}
\end{equation}
where ${\bf T}_{2}=\tanh^{2}\beta J$, 
\begin{displaymath}
K{\bf T}_{2,c}^{SG}=\frac{1}{\frac{\left(1-\mu \right)^2}{1-2\mu} +1}= 
\frac{1}{\frac{\left(\gamma-2 \right)^2}{\left( \gamma-1\right)\left(\gamma-3\right)} +1} = K \tanh^{2} \frac{J}{T_{c}^{SG}}
\end{displaymath}
(see Eq.\ (\ref{TcSG1eq})), $Q=\left( 1-\mu\right) K{\bf T}_{2}q$ and 
$D\left( \lambda\right) =\frac{1}{2} I_{1}\left( \lambda, 0\right)$ $= \int_{0}^{\infty} x^{3-2\lambda} \left(\tanh^{2}x-x^{2}\right) dx <0$.
From Eq.\ (\ref{qAqBsg1eq}) follows that $q\propto \varepsilon^{1/\left(\gamma -3\right)}$ just below the transition point. 
This is the same scaling relation as in 
the case of the SG transition in the Ising model on SF networks \cite{Kim05}, only the critical temperature is different.
For $\gamma>4$ the following equation for $q$ is obtained from Eq.\ (\ref{qAsg3}),
\begin{equation}
\left( \frac{1}{K{\bf T}_{2}}-\frac{1}{K {\bf T}_{2,c}^{SG}} \right) Q =
2 \left[ \frac{\left( \gamma-2\right)^{2}}{\left(4- \gamma\right)\left(\gamma-1\right)}
+3 \frac{\left( \gamma -2\right)}{3-\gamma}\right]Q^{2}.
\label{qAqBsg3eq}
\end{equation}
Hence, $q\propto \varepsilon$ just below the SG transition temperature,  as in 
the case of the SG transition in the Ising model on SF networks \cite{Kim05}; only the critical temperature and the proportionality
constant are different. Thus, it can be seen that the critical exponent for the SG order parameter exhibits the same scaling behavior
in the case of the SG transition in the Ising model on a SF networks and on a MN with SF layers with identical distributions of the degrees of nodes. 

If the SF layers have different distributions of the degrees of nodes the critical exponents for $q^{(A)}$, $q^{(B)}$ must be determined
from two-dimensional systems of nonlinear equations following from Eq.\ (\ref{qAsg3}) and from
the complementary equation for $Q^{(B)}/K^{(B)}{\bf T}_{2}^{(B)}$ in which only the lowest-order nonlinear terms are retained.
Due to complexity of these equations as well as the complex form of $T_{c}^{SG}$, Eq.\ (\ref{TcSG1}), it is not a simple task to determine
the scaling behavior of $q^{(A)}$, $q^{(B)}$ analytically. Hence, below this is done by solving the above-mentioned systems 
numerically for $q^{(A)}$, $q^{(B)}$ vs.\ $\varepsilon$. 

For $3 <\gamma^{(A)} <4$, $3< \gamma^{(B)} <4$ the following
system of equations for $q^{(A)}$, $q^{(B)}$ is obtained from Eq.\ (\ref{qAsg3}),
\begin{eqnarray}
\frac{Q^{(A)}}{K^{(A)}{\bf T}_{2}^{(A)}} &=& \frac{\left( \gamma^{(A)}-2\right)^{2}}{\left( \gamma^{(A)}-1\right)}
2^{\gamma^{(A)}-1} \frac{\Gamma\left( \gamma^{(A)}-\frac{3}{2}\right)}{\sqrt{\pi}} D\left(\gamma^{(A)}\right) 
\left( Q^{(A)}\right)^{\gamma^{(A)}-2}
\nonumber\\
&+&  \frac{\left( \gamma^{(A)}-2\right)^{2}}{\left( \gamma^{(A)}-1\right)\left( \gamma^{(A)}-3\right)} Q^{(A)} 
+ \frac{\left( \gamma^{(A)}-2\right)}{\left( \gamma^{(A)}-1\right)} \frac{\left( \gamma^{(B)}-1\right)}{\left( \gamma^{(B)}-2\right)} Q^{(B)},
\nonumber\\
\frac{Q^{(B)}}{K^{(B)}{\bf T}_{2}^{(B)}} &=& \frac{\left( \gamma^{(B)}-2\right)^{2}}{\left( \gamma^{(B)}-1\right)}
2^{\gamma^{(B)}-1} \frac{\Gamma\left( \gamma^{(B)}-\frac{3}{2}\right)}{\sqrt{\pi}} D\left(\gamma^{(B)}\right) 
\left( Q^{(B)}\right)^{\gamma^{(B)}-2}
\nonumber\\
&+&  \frac{\left( \gamma^{(B)}-2\right)}{\left( \gamma^{(B)}-1\right)} \frac{\left( \gamma^{(A)}-1\right)}{\left( \gamma^{(A)}-2\right)} Q^{(A)}
+ \frac{\left( \gamma^{(B)}-2\right)^{2}}{\left( \gamma^{(B)}-1\right)\left( \gamma^{(B)}-3\right)} Q^{(B)}.
\label{qAqBsg1}
\end{eqnarray}
It is then possible, e.g., to evaluate $Q^{(B)}$ from the first equation and insert it to the second one to obtain a nonlinear equation for $Q^{(A)}$
which can be solved numerically. The results of this procedure are summarized in Fig.\ 3. For $\gamma^{(A)} =\gamma^{(B)}=3.5$ 
($\mu^{(A)}=\mu^{(B)}=1/2$) the scaling relation $q^{(A)}\propto \varepsilon^{\beta}$ is obtained with $\beta =1.98$ which agrees well
with $\beta=1/\left( \gamma^{(A)}-3\right) =2$ predicted from Eq.\ (\ref{qAqBsg1eq}). For $\gamma^{(A)}=3.5$ and
$4> \gamma^{(B)}=3.75 > \gamma^{(A)}$ ($\mu^{(B)}=4/11$)  the scaling relation $q^{(A)}\propto \varepsilon^{\beta}$ is obtained with $\beta =1.95$
which is still close to $\beta = 1/\left(\gamma^{(A)} -3\right) =2$ (the difference can be due to fitting error); thus, the scaling behavior of $q^{(A)}$
in the vicinity of $T_{c}^{SG}$ is still determined by the lowest-order nonlineatity $\left( Q^{(A)}\right)^{\gamma^{(A)}-2}$ in Eq.\ (\ref{qAqBsg1}). 
Since in the first approximation $Q^{(B)}\propto Q^{(A)}$ the scaling behavior of $q^{(B)}$ is the same as that of $q^{(A)}$.

For $3 <\gamma^{(A)} <4$, $\gamma^{(B)} >4$ in the system of equations for $q^{(A)}$, $q^{(B)}$ obtained from Eq.\ (\ref{qAsg3})
the equation for $Q^{(A)}/K^{(A)}{\bf T}_{2}^{(A)}$ is identical as in Eq.\ (\ref{qAqBsg1}) and the second equation is
\begin{eqnarray}
\frac{Q^{(B)}}{K^{(B)}{\bf T}_{2}^{(B)}} &=&
 \frac{\left( \gamma^{(B)}-2\right)}{\left( \gamma^{(B)}-1\right)} \frac{\left( \gamma^{(A)}-1\right)}{\left( \gamma^{(A)}-2\right)} Q^{(A)}
+ \frac{\left( \gamma^{(B)}-2\right)^{2}}{\left( \gamma^{(B)}-1\right)\left( \gamma^{(B)}-3\right)} Q^{(B)}
\nonumber\\
&-&\frac{\left( \gamma^{(B)}-2\right)^{2}\left( \gamma^{(A)}-1\right)}{\left( \gamma^{(B)}-1\right)}
\left[ \frac{2Q^{(A)2}}{\left( \gamma^{(A)}-3\right)\left( \gamma^{(B)}-2\right)}  \right. \nonumber\\
&+& \left. \frac{4 Q^{(A)}Q^{(B)}}{\left( \gamma^{(A)}-2\right)\left( \gamma^{(B)}-3\right)}
+ \frac{2Q^{(B)2}}{\left( \gamma^{(A)}-1\right)\left( \gamma^{(B)}-4\right)} \right].
\label{qAqBsg2}
\end{eqnarray}
It is again possible to evaluate $Q^{(B)}$ from the first equation in Eq.\ (\ref{qAqBsg1}) and insert it into Eq.\ (\ref{qAqBsg2}) 
to obtain a nonlinear equation for $Q^{(A)}$ which can be solved numerically. 
For $\gamma^{(A)}=3.5$ and
$\gamma^{(B)}=4.5 > 4$ ($\mu^{(B)}=2/7$)  the scaling relation $q^{(A)}\propto \varepsilon^{\beta}$ is obtained with $\beta =2.0$ (Fig.\ 3),
again in agreement with $\beta = 1/\left( \gamma^{(A)}-3 \right) =2$; thus, the scaling behavior of $q^{(A)}$
in the vicinity of $T_{c}^{SG}$ is again determined by the lowest-order nonlineatity $\left( Q^{(A)}\right)^{\gamma^{(A)}-2}$ in 
the first equation in Eq.\ (\ref{qAqBsg1}).

For $\gamma^{(A)}>4$, $\gamma^{(B)}>4$, 
in the system of equations for $q^{(A)}$, $q^{(B)}$ obtained from Eq.\ (\ref{qAsg3})
the equation for $Q^{(B)}/K^{(B)}{\bf T}_{2}^{(B)}$ is identical as in Eq.\ (\ref{qAqBsg2}) and the equation 
for $Q^{(A)}/K^{(A)}{\bf T}_{2}^{(A)}$ can be obtained from it by replacing $(A)$ with $(B)$ and vice versa. The system of 
two quadratic equations for $Q^{(A)}$, $Q^{(B)}$ can be solved by evaluating, e.g., $Q^{(A)}$ from one equation,
inserting it to the other one and solving for $Q^{(B)}$ numerically. From such solution, independently of $\gamma^{(A)}$, $\gamma^{(B)}$,
the scaling relation $q^{(A)} \propto \varepsilon$ is obtained with high accuracy, i.e., the scaling exponent $\beta =1$ is the same as that
resulting from Eq.\ (\ref{qAqBsg3eq}) valid for $\gamma^{(A)}=\gamma^{(B)}$. Thus, the scaling behavior of $q^{(A)}$
in the vicinity of $T_{c}^{SG}$ is again determined by the lowest-order nonlineatity $\left( Q^{(A)}\right)^{2}$ in Eq.\ (\ref{qAqBsg2}).

It is interesting to note that for $\gamma^{(A)}>3$, $\gamma^{(B)}>3$, i.e., when $T_{c}^{SG}$ is finite 
the last term in Eq.\ (\ref{qAsg3}) containing $\left( Q^{(A)}\right)^{\gamma^{(A)}-2} \left( Q^{(B)}\right)^{\gamma^{(B)}-1}$ 
is of higher order than the leading nonlinear terms in Eq.\ (\ref{qAqBsg1}) and Eq.\ (\ref{qAqBsg2}) and can be omitted.
Let us also mention that taking into account the results of Ref.\ \cite{Kim05} logarithmic corrections to the scaling behavior
of $q^{(A)}$, $q^{(B)}$ in the vicinty of $T_{c}^{SG}$ can be expected for $\gamma^{(A)}$, $\gamma^{(B)}$ being integer 
numbers. Unfortunately, verification of the presence of such correction from numerical solution of the systems of nonlinear equations like
Eq.\ (\ref{qAqBsg1}) or Eq.\ (\ref{qAqBsg2}) is practically impossible, thus this case is not considered in this paper.

\section{Summary and conclusions}

In this paper the SG and FM transitions in the Ising model on MNs with both FM and AFM interactions were studied using the
replica method; in particular, the MNs with the layers in the form of random ER graphs and SF networks were considered. The critical
temperatures for these transitions from the paramagnetic phase were determined from the RS solution. For the Ising model on MNs with
SF layers it was shown that the transition temperature is finite if the distributions of the degrees of nodes within both layers have a finite
second moment, and that depending on the model parameters this transition can be to the FM or SG phase. It was also shown that the 
correlation between the degrees of nodes within different layers significantly influences the critical temperatures for the FM and SG transitions
and thus the phase diagram. In particular, in the case of MN with two independently generated SF layers corresponding to (balanced) FM and 
AFM interactions for small mean degrees of nodes within both layers the transition is to the FM or SG phase, depending on
the details of the two degree distributions, while for high mean degrees of nodes it is to the FM phase; this result was confirmed in MC simulations.
In contrast, in the case of MNs with maximally correlated layers the transition can be to the SG phase also for high mean degrees of nodes
within the two layers, depending on the details of the two degree distributions. The scaling behavior for the SG parameter was determined
from the RS solution by means of a semi-analytic procedure. In most cases the critical exponent has the universal value $\beta =1$, only in the
case of SF layers characterized by the distributions of the degrees of nodes with diverging third moment its value becomes dependent on the
details of this distribution.

Application of the replica method to the Ising model on MNs with both FM and AFM interactions leads to many results which
have already been reported for the Ising model on MNs with purely FM interactions, e.g., the occurrence, in a natural way, of the
sets of the order paramaters associated with consecutive layers or the dependence of the critical temperature on the correlations
between the degrees of nodes within different layers \cite{Krawiecki17}. Although the methods used in this paper are direct generalization
of the ones used in Refs.\ \cite{Viana85,Kim05} to the case of MNs with possibly heterogeneous layers both the calculations and the 
results exhibit some peculiar properties due to the fact that the statistical averages must be evaluated over the separately generated
layers rather than over the whole network. The problems remaining for future research comprise, e.g., 
investigation of the stability of the FM and SG phases, obtaining 
the full phase diagram for the Ising model on MNs and determination of the critical behavior of the model at the FM and SG phase borders.

\section*{Appendix A}

In networks generated from the static model there is $\langle k\rangle =K$, 
$N\sum_{i} v_{i}^{2} =\left( \langle k^{2}\rangle -\langle k\rangle\right)/\langle k\rangle^2$ \cite{Lee04}. Thus the result of 
Eq.\ (\ref{TcFM1}) and  Eq.\ (\ref{TcSG1}) can be written in a more general form,
\begin{equation}
T_{c}^{FM}=J {\rm atanh}^{-1} \left\{
\frac{\frac{\langle k^{(A)2}\rangle}{\langle k^{(A)} \rangle} - \frac{\langle k^{(B)2}\rangle}{\langle k^{(B)} \rangle} -\sqrt{\Delta}}
{2 \left[ \langle k^{(A)} \rangle \langle k^{(B)} \rangle -
\left( \frac{\langle k^{(A)2}\rangle}{\langle k^{(A)} \rangle}-1\right) \left( \frac{\langle k^{(B)2}\rangle}{\langle k^{(B)} \rangle} -1\right)
\right]}
\right\}
\label{TcFM1gen}
\end{equation}
where
\begin{displaymath}
\Delta = \left( \frac{\langle k^{(A)2}\rangle}{\langle k^{(A)} \rangle} - \frac{\langle k^{(B)2}\rangle}{\langle k^{(B)} \rangle} -2 \right)^{2}
-4 \langle k^{(A)} \rangle \langle k^{(B)} \rangle,
\end{displaymath}
and
\begin{equation}
T_{c}^{SG}=J {\rm atanh}^{-1} \left\{
\frac{\frac{\langle k^{(A)2}\rangle}{\langle k^{(A)} \rangle} + \frac{\langle k^{(B)2}\rangle}{\langle k^{(B)} \rangle} -2 -\sqrt{\Delta}}
{2 \left[\left( \frac{\langle k^{(A)2}\rangle}{\langle k^{(A)} \rangle}-1\right) \left( \frac{\langle k^{(B)2}\rangle}{\langle k^{(B)} \rangle} -1\right)
- \langle k^{(A)} \rangle \langle k^{(B)} \rangle \right]}
\right\}^{1/2}
\label{TcSG1gen}
\end{equation}
where
\begin{displaymath}
\Delta = \left( \frac{\langle k^{(A)2}\rangle}{\langle k^{(A)} \rangle} - \frac{\langle k^{(B)2}\rangle}{\langle k^{(B)} \rangle} \right)^{2}
+4 \langle k^{(A)} \rangle \langle k^{(B)} \rangle,
\end{displaymath}
using the moments of the distributions of the degrees of nodes within each layer. 

It should be noted that the necessary condition for the occurrence of the FM or SG transition is that the critical temperatures $T_{c}^{FM}$,
$T_{c}^{SG}$ given by Eq.\ (\ref{TcFM1gen}) and Eq.\ (\ref{TcSG1gen}), respectively, are real and positive. 
In the case of the critical temperature for the SG transition it can be easily shown that this requires that
\begin{equation}
\frac{\langle k^{(A)2}\rangle}{\langle k^{(A)} \rangle} + \frac{\langle k^{(B)2}\rangle}{\langle k^{(B)} \rangle} +\sqrt{\Delta} >4,
\end{equation}
which is also a condition for the occurrence of a giant component  in a MN with two independently generated layers 
\cite{Lee14}, in which nodes are connected via edges in any layer (but not necessarily in both layers). Thus, in the case of a MN with 
two layers, one with purely FM and the other one with purely AFM interactions, the SG transition can appear
in a MN above the percolation threshold, in analogy with the case of complex networks \cite{Kim05}. In contrast, this is not enough for the
possibility of appearance of the FM transition, for which higher densities of connections within layers are necessary.

It can be expected that Eq.\ (\ref{TcFM1gen}) and Eq.\ (\ref{TcSG1gen}) are valid for any multiplex
network consisting of independently generated, possibly heterogeneous
layers with finite second moments of the distributions of the degrees of nodes (note, however, that in the case of random ER layers
the above expressions are not properly determined; this is since the equations for $T_{c}^{FM}$ ($T_{c}^{SG}$) are linear (quadratic)
rather than quadratic (quartic) with respect to $\tanh \beta J$, see Sec.\ 4.2).
In particular, if the distributions of the degrees of nodes within the layers $G^{(A)}$, $G^{(B)}$ obey power scaling laws in the form
$p_{k^{(A)}} = \left( \gamma^{(A)}-1\right) \left( \tilde{m}^{(A)}\right)^{\gamma^{(A)}-1} \left( k^{(A)}\right)^{-\gamma^{(A)}}$ for 
$k^{(A)}>\tilde{m}^{(A)}$, 
$p_{k^{(B)}} = \left( \gamma^{(B)}-1\right) \left( \tilde{m}^{(B)}\right)^{\gamma^{(B)}-1} \left( k^{(B)}\right)^{-\gamma^{(B)}}$ for 
$k^{(B)}>\tilde{m}^{(B)}$, respectively,
with fixed minimum degrees of nodes $\tilde{m}^{(A)}$, $\tilde{m}^{(B)}$
the critical temperature can be obtained by inserting in  Eq.\ (\ref{TcFM1gen}) and Eq.\ (\ref{TcSG1gen})
\begin{displaymath}
\frac{\langle k^{(A)2}\rangle}{\langle k^{(A)} \rangle}= \tilde{m}^{(A)} \frac{\gamma^{(A)}-2}{\gamma^{(A)}-3}, \;\; 
\langle k^{(A)} \rangle = \tilde{m}^{(A)} \frac{\gamma^{(A)}-1}{\gamma^{(A)}-2},
\end{displaymath}
and similar expressions for the moments of $p_{k^{(B)}}$. 

\section*{Appendix B}

In this Appendix a general expansion formula is derived for a sum of a form
\begin{equation}
S\left( y_{A},y_{B}\right) = \left( \lambda_{A}-1\right) \left( \lambda_{B}-1\right) y_{A}^{\lambda_{A}-2} y_{B}^{\lambda_{B}-1} 
\int_{y_{A}}^{\infty} \int_{y_{B}}^{\infty} \frac{F\left( x_{1},x_{2}\right) }{x_{1}^{\lambda_{A}} x_{2}^{\lambda_{B}}} dx_{2} dx_{1},
\label{s}
\end{equation}
using a method which is a generalization of the one from Ref.\ \cite{Kim05} to the case of two-dimensional integrals. 

Let us assume that $\lambda_{A}$, $\lambda_{B}$ are not integer numbers and, for some $m_{1}$, $m_{2}$, there is $m_{1} < \lambda_{A} < m_{1}+1$,
$m_{2} < \lambda_{B} < m_{2} +1$. The expansion of the function $F\left( x_{1},x_{2}\right)$ in the Taylor series is
\begin{eqnarray}
F\left( x_{1},x_{2}\right) &= & \sum_{n_{1},n_{2}=0}^{\infty} f_{n_{1},n_{2}} x_{1}^{n_{1}}x_{2}^{n_{2}}=  \nonumber \\
&=& \sum_{n_{1}=0}^{m_{1}-1} \sum_{n_{2}=0}^{m_{2}-1} f_{n_{1},n_{2}} x_{1}^{n_{1}}x_{2}^{n_{2}} \nonumber \\
&+&  \sum_{n_{1}=m_{1}}^{\infty} \sum_{n_{2}=0}^{m_{2}-1} f_{n_{1},n_{2}} x_{1}^{n_{1}}x_{2}^{n_{2}}
+ \sum_{n_{1}=0}^{m_{1}-1} \sum_{n_{2}=m_{2}}^{\infty} f_{n_{1},n_{2}} x_{1}^{n_{1}}x_{2}^{n_{2}} \nonumber \\
&+&  \sum_{n_{1}=m_{1}}^{\infty} \sum_{n_{2}=m_{2}}^{\infty} f_{n_{1},n_{2}} x_{1}^{n_{1}}x_{2}^{n_{2}},
\label{Taylorexp}
\end{eqnarray}
where the expansion coefficients are
\begin{equation}
f_{n_{1},n_{2}} = \frac{1}{\left( n_{1}+n_{2}\right) !} {n_{1}+n_{2} \choose n_{1}} 
\left. \frac{\partial ^{n_{1}+n_{2}} F}{\partial x_{1}^{n_{1}} \partial x_{2}^{n_{2}}} \left( x_{1},x_{2}\right) \right|_{(0,0)}.
\label{Taylorcoeff}
\end{equation}
The first sum in Eq.\ (\ref{Taylorexp}) can be integrated term by term which yields
\begin{eqnarray}
&& \sum_{n_{1}=0}^{m_{1}-1} \sum_{n_{2}=0}^{m_{2}-1} f_{n_{1},n_{2}} 
\int_{y_{A}}^{\infty} x_{1}^{n_{1}-\lambda_{A}} dx_{1} \int_{y_{B}}^{\infty} x_{2}^{n_{2}-\lambda_{B}} dx_{2} \nonumber \\
&& = \sum_{n_{1}=0}^{m_{1}-1} \sum_{n_{2}=0}^{m_{2}-1} f_{n_{1},n_{2}}
\frac{(-1)}{n_{1}-\lambda_{A}+1} \frac{(-1)}{n_{2}-\lambda_{B}+1} y_{A}^{n_{1}-\lambda_{A}+1} y_{B}^{n_{2}-\lambda_{B}+1}.
\label{c1}
\end{eqnarray}

Concerning the remaining terms it should be noted that in the converging Taylor series, Eq.\ (\ref{Taylorexp}), the order of summation and
integration can be exchanged. Then, e.g., from integration of the second sum, after evaluating the integral
$\int_{y_{B}}^ {\infty} x_{2}^{n_{2}-\lambda_{B}} dx_{2}= - \left( n_{2}-\lambda_{B} +1 \right)^{-1} y_{B}^{n_{2}-\lambda_{B} +1}$
for $0 \le n_{2} \le m_{2}-1$ and dividing in two parts and evaluating the integral
$\int_{y_{A}}^ {\infty} x_{1}^{n_{1}-\lambda_{A}} dx_{1}=\left( \int_{0}^ {\infty} - \int_{0}^ {y_{A}} \right) x_{1}^{n_{1}-\lambda_{A}} dx_{1}=
\int_{0}^ {\infty}x_{1}^{n_{1}-\lambda_{A}} dx_{1} - \left( n_{1}-\lambda_{A} +1 \right)^{-1} y_{A}^{n_{1}-\lambda_{A} +1}$
for $n_{1} \ge m_{1}$ it is obtained that
\begin{eqnarray}
&& \sum_{n_{1}=m_{1}}^{\infty} \sum_{n_{2}=0}^{m_{2}-1} f_{n_{1},n_{2}} 
\int_{y_{A}}^ {\infty} x_{1}^{n_{1}-\lambda_{A}} dx_{1} \int_{y_{B}}^ {\infty} x_{2}^{n_{2}-\lambda_{B}} dx_{2} = \nonumber \\
&& - \sum_{n_{2}=0}^{m_{2}-1} I_{1}\left( \lambda_{A},n_{2}\right) 
\frac{y_{B}^{ n_{2}-\lambda_{B} +1}}{ n_{2}-\lambda_{B} +1} 
+ \sum_{n_{2}=0}^{m_{2}-1} \sum_{n_{1}=m_{1}}^{\infty} f_{n_{1},n_{2}} 
\frac{y_{A}^{ n_{1}-\lambda_{A} +1}}{ n_{1}-\lambda_{A} +1} \frac{y_{B}^{ n_{2}-\lambda_{B} +1}}{ n_{2}-\lambda_{B} +1}, \nonumber\\
&& \label{c2}
\end{eqnarray}
where
\begin{equation}
 I_{1}\left( \lambda_{A},n_{2}\right) = \sum_{n_{1}=m_{1}} ^{\infty}f_{n_{1},n_{2}} \int_{0}^{\infty} x_{1}^{n_{1}-\lambda_{A}} dx_{1}.
\label{I1}
\end{equation}
Since $m_{1}< \lambda_{A}<m_{1}+1$ the integrals in Eq.\ (\ref{I1}) are not singular and the whole series converges. Taking into account that, from
Eq.\ (\ref{Taylorcoeff}),
\begin{eqnarray}
\sum_{n_{1}=0}^{\infty} f_{n_{1},n_{2}}x_{1}^{n_{1}}&=& \frac{1}{n_{2}!} \sum_{n_{1}=0}^{\infty} \frac{1}{n_{1}!} 
\frac{\partial^{n_{1}}}{\partial x_{1}^{n_{1}}} \left.  \left[ \frac{\partial^{n_{2}}F}{\partial x_{2}^{n_{2}}} \left( x_{1},x_{2} \right)\right] \right|_{(0,0)}
x_{1}^{n_{1}} = \frac{1}{n_{2}!} \left. \frac{\partial^{n_{2}}F}{\partial x_{2}^{n_{2}}} \left( x_{1},x_{2} \right) \right|_{x_{2}=0}, \nonumber\\
&&
\end{eqnarray}
Eq.\ (\ref{I1}) can be rewritten as
\begin{eqnarray}
 I_{1}\left( \lambda_{A},n_{2}\right) &= & \int_{0}^{\infty} x_{1}^{-\lambda_{A}}
\left( \sum_{n_{1}=0}^{\infty} f_{n_{1},n_{2}}x_{1}^{n_{1}} - \sum_{n_{1}=0}^{m_{1}-1} f_{n_{1},n_{2}}x_{1}^{n_{1}} \right) dx_{1} 
\nonumber\\
&=& \int_{0}^{\infty} x_{1}^{-\lambda_{A}}
\left(  \frac{1}{n_{2}!} \left. \frac{\partial^{n_{2}}F}{\partial x_{2}^{n_{2}}} \left( x_{1},x_{2} \right) \right|_{x_{2}=0}
- \sum_{n_{1}=0}^{m_{1}-1} f_{n_{1},n_{2}}x_{1}^{n_{1}} \right) dx_{1}. \nonumber\\
&& \label{I11}
\end{eqnarray}
The third and fourth sum in Eq.\ (\ref{Taylorexp}) can be integrated in a similar way, and finally from Eq.\ (\ref{s}) it is obtained that
\begin{eqnarray}
S\left( y_{A},y_{B}\right) &=&\left( \lambda_{A}-1\right) \left( \lambda_{B}-1\right) \left[
\sum_{n_{1}=0}^{\infty} \sum_{n_{2}=0}^{\infty} 
f_{n_{1},n_{2}}\frac{y_{A}^{n_{1}-1} y_{B}^{n_{2}}}{\left( n_{1}-\lambda_{A}+1\right) \left( n_{2}-\lambda_{B}+1\right)}
\right. \nonumber\\
&&  - \sum_{n_{1}=0}^{\infty} I_{2}\left( \lambda_{B},n_{1}\right) 
\frac{y_{A}^{ n_{1}-1} y_{B}^{\lambda_{B}-1}}{ n_{1}-\lambda_{A} +1} 
 - \sum_{n_{2}=0}^{\infty} I_{1}\left( \lambda_{A},n_{2}\right) 
\frac{y_{A}^{\lambda_{A}-2}y_{B}^{n_{2}}}{ n_{2}-\lambda_{B} +1} \nonumber\\
&& \left. +  I\left( \lambda_{A}, \lambda_{B}\right) y_{A}^{\lambda_{A}-2}y_{B}^{\lambda_{B}-1} \right],
\label{Sfinal}
\end{eqnarray}
where
\begin{eqnarray}
I_{2}\left( \lambda_{B},n_{1}\right)&=& \sum_{n_{2}=m_{2}} ^{\infty}f_{n_{1},n_{2}} \int_{0}^{\infty} x_{2}^{n_{2}-\lambda_{B}} dx_{2}
\nonumber\\
&=& \int_{0}^{\infty} x_{2}^{-\lambda_{B}}
\left(  \frac{1}{n_{1}!} \left. \frac{\partial^{n_{1}}F}{\partial x_{1}^{n_{1}}} \left( x_{1},x_{2} \right) \right|_{x_{1}=0}
- \sum_{n_{2}=0}^{m_{2}-1} f_{n_{1},n_{2}}x_{2}^{n_{2}} \right) dx_{2}, \nonumber\\
\label{I2}
&&
\end{eqnarray}
\begin{eqnarray}
 I\left( \lambda_{A}, \lambda_{B}\right) &=&
\sum_{n_{1}=m_{1}}^{\infty} \sum_{n_{2}=m_{2}}^{\infty} f_{n_{1},n_{2}} 
\int_{y_{A}}^{\infty} x_{1}^{n_{1}-\lambda_{A}} dx_{1} \int_{y_{B}}^{\infty} x_{2}^{n_{2}-\lambda_{B}} dx_{2} \nonumber \\
&=& \int_{0}^{\infty}\int_{0}^{\infty} x_{1}^{-\lambda_{A}}x_{2}^{-\lambda_{B}} \left[ F\left( x_{1},x_{2}\right) -
\sum_{n_{2}=0}^{m_{2}-1} \frac{x_{2}^{n_{2}}}{n_{2}!}  \left. \frac{\partial^{n_{2}}F}{\partial x_{2}^{n_{2}}} \left( x_{1},x_{2} \right) \right|_{x_{2}=0}
\right. \nonumber \\
&& \left.
- \sum_{n_{1}=0}^{m_{1}-1}  
 \frac{x_{1}^{n_{1}}}{n_{1}!} \left. \frac{\partial^{n_{1}}F}{\partial x_{1}^{n_{1}}} \left( x_{1},x_{2} \right) \right|_{x_{1}=0}
+ \sum_{n_{1}=0}^{m_{1}-1} \sum_{n_{2}=0}^{m_{2}-1} f_{n_{1},n_{2}} x_{1}^{n_{1}}x_{2}^{n_{2}} \right] dx_{1}dx_{2}
\nonumber\\
\label{I}
\end{eqnarray}
and the integrals in Eq.\ (\ref{I2}) and Eq.\ (\ref{I}) converge.

In particular, it can be seen that for $F\left( x_{1},x_{2}\right) = x_{1}\tanh^{2} \sqrt{x_{1}+x_{2}}$ there is
$f_{0,0}=f_{1,0}=0$ and $f_{0,n_{2}}=0$ for $n_{2}=0,1,2\ldots$,
thus also $I_{2}\left( \lambda_{B},0\right)=0$ from Eq.\ (\ref{I2}) and terms corresponding to $n_{1}=0$ (containing $y_{A}^{-1}$)
in the sums in Eq.\ (\ref{Sfinal}) disappear.

\end{document}